\documentclass[useAMS,usenatbib]{mn2e}

\usepackage{times,graphicx}
\usepackage[total={17.8cm,24.0cm},centering]{geometry}

\newcommand{\farcsec}{\hbox{$.\!\!^{\prime\prime}$}}  	
\newcommand{\kms}{km~s$^{-1}$}
\newcommand{\mbh}{M$_{\rm BH}$} 				
\newcommand{\msun}{M$_{\odot}$}					
\newcommand{\ml}{M$_{\odot}$/L$_{\odot}$}			
\newcommand{\bhs}{\mbh-$\sigma$}				

\newcommand{\ha}{H$\alpha$}					
\newcommand{\hb}{H$\beta$}						
\newcommand{\mgb}{Mg$\,b$}						
\newcommand{\nidoub}{[N{\small I}]}				
\newcommand{\nii}{[N{\small II}]}					
\newcommand{\oiii}{[O{\small III}]}					
\newcommand{\han}{\ha+\nii}						

\newcommand{\rbh}{$0\farcsec29$}					
\newcommand{\sbh}{$1.4 \times 10^8$ \msun}	 		
\newcommand{\sbhedgeerr}{$1.4^{+2.6}_{-0.2} \times 10^8$ \msun}    
\newcommand{\sbherr}{$1.4^{+2.6}_{-1.0} \times 10^8$ \msun}    
\newcommand{\sbhone}{$1.4^{+0.3}_{-0.8} \times 10^8$ \msun}    
\newcommand{\gbh}{$2.0 \times 10^8$ \msun}	 		
\newcommand{\gbherr}{$(2.0 \pm 0.1) \times 10^8$ \msun}    

\newcommand{\sml}{3.08}					
\newcommand{\smlerr}{$3.08 \pm 0.2$}			
\newcommand{\gmlerr}{$3.0 \pm 0.4$}			

\title[The Black Hole in NGC 3379]{The Black Hole in NGC 3379: A Comparison of Gas and Stellar Dynamical Mass Measurements with {\it HST} and Integral-Field Data\thanks{Based on observations obtained with the {\it Hubble Space Telescope}, which is operated by AURA, Inc., under NASA contract NAS5-2655.  Also based on observations obtained at the Canada-France-Hawaii Telescope (CHFT) which is operated by the National Research Council of Canada, the Institut National des Sciences de l'Univers of the Centre National de la Recherche Scientifique of France, and the University of Hawaii.}}
\author[K. L. Shapiro et al.]{Kristen L. Shapiro,$^{1,2}$\thanks{E-mail: shapiro@astro.berkeley.edu} Michele Cappellari,$^2$ Tim de Zeeuw,$^2$ Richard M. McDermid,$^2$
\newauthor
Karl Gebhardt,$^3$ Remco C. E. van den Bosch,$^2$ and Thomas S. Statler$^{4}$ \\
$^{1}$UC Berkeley Department of Astronomy, Berkeley, CA 94720, USA\\
$^{2}$Leiden Observatory, Postbus 9513, 2300 RA Leiden, The Netherlands\\
$^{3}$UT Austin Department of Astronomy, Austin, TX 78712, USA\\
$^{4}$Ohio University Department of Physics and Astronomy, Athens, OH 45701, USA} 

\begin{document}

\date{}

\pagerange{\pageref{firstpage}--\pageref{lastpage}} \pubyear{}

\maketitle

\label{firstpage}

\begin{abstract}
We combine {\it Hubble Space Telescope} spectroscopy and ground-based integral-field data from the {\tt SAURON} and {\tt OASIS} instruments to study the central black hole in the nearby elliptical galaxy NGC~3379.  From these data, we obtain kinematics of both the stars and the nuclear gaseous component.  Axisymmetric three-integral models of the stellar kinematics find a black hole of mass \sbherr\ (3$\sigma$ errors).  These models also probe the velocity distribution in the immediate vicinity of the black hole and reveal a nearly isotropic velocity distribution throughout the galaxy and down to the black hole sphere of influence $R_{\rm BH}$.  The morphology of the nuclear gas disc suggests that it is not in the equatorial plane; however the core of NGC~3379 is nearly spherical.  Inclined thin-disc models of the gas find a nominal black hole of mass \gbherr\ (3$\sigma$ errors), but the model is a poor fit to the kinematics.  The data are better fit by introducing a twist in the gas kinematics (with the black hole mass assumed to be \gbh), although the constraints on the nature and shape of this perturbation are insufficient for more detailed modelling.  Given the apparent regularity of the gas disc's appearance, the presence of such strong non-circular motion indicates that caution must be used when measuring black hole masses with gas dynamical methods alone.
\end{abstract}


\section{Introduction}

In the study of galaxy formation and evolution, one of the more intriguing recent results is the so-called black hole - sigma (\bhs) relation between the mass of a galaxy's central supermassive black hole (SMBH) and the large-scale velocity dispersion of the stars (e.g., Ferrarese \& Merritt 2000; Gebhardt et al. 2000a).  What is particularly remarkable is the tightness of this correlation, which spans three orders of magnitude in black hole mass and one in stellar velocity dispersion (but see Bender et al. 2005).  The primary implication of the existence of the \bhs\ relation is clear: over a large range of galaxy sizes, there must be a fundamental connection between the evolution of the very small central regions of an individual galaxy to that of the galaxy on much larger scales.  This connection has yet to be completely understood; however, it is thought to reflect the SMBHs regulating the amount of gas and thus the rate of star formation in their host galaxies' spheroids (e.g. Silk \& Rees 1998; Granato et al. 2004).  Understanding SMBHs and their relation to their environments can therefore provide much insight into the formation and merger history of their host galaxies.

In practice, the \bhs\ result is the product of nearly 40 measurements of SMBH masses in nearby galaxies using several distinct techniques (see e.g. Kormendy \& Gebhardt 2001 and Barth 2004 for reviews).  By far the most reliable of these measurements is that of the SMBH in Milky Way, the relative proximity of which allows the orbits of individual stars in the vicinity of the black hole to be traced in great detail (Eckart \& Genzel 1997; Genzel et al. 1997, 2000; Ghez et al. 1998, 2000).  As to extragalactic BHs, the most accurate measurements are studies of H$_2$O maser emission in galaxy nuclei, which have been observed to follow nearly perfect Keplerian rotation around a central massive object (e.g., Miyoshi et al. 1995; Herrnstein et al. 2005).  Although quite straightforward and precise, this method relies on a galaxy having detectable nuclear maser emission and therefore accounts for only a handful of the SMBH mass measurements.  A similar method takes advantage of galaxies with nuclear ionised gas discs; assuming the gas is contained in a thin disc and is influenced solely by the combined gravitational potential of the galaxy and a central black hole, the kinematics of these discs can be used to estimate the mass of the black hole (e.g., Harms et al. 1994; Macchetto et al. 1997; van der Marel \& van den Bosch 1998; Bertola et al. 1998; Barth et al. 2001).  This technique has also recently been applied to the nuclear stellar disc in M31, which is unique in being sufficiently resolved for the stellar Keplerian motion to be measured (Bender et al. 2005).  The most used technique for measuring SMBH mass, however, relies not on detecting Keplerian motion but rather on detailed models of the stellar dynamics.  Over two-thirds of SMBH mass measurements are the results of long-slit stellar kinematic data, coupled with axisymmetric dynamical models (e.g., van der Marel et al. 1998; Cappellari et al. 2002; Gebhardt et al. 2003).

These gas dynamical and stellar dynamical methods, despite being responsible for nearly all black hole mass estimates, are often limited both by the necessary assumptions and by other factors such as spatial resolution of the observations and computational resources.  Despite such uncertainties, measurements with these various techniques nevertheless result in a tight \bhs\ correlation, and this is compelling evidence that the relationship is genuine.  However, these methods are expected to suffer mainly from systematic effects, mostly influencing the slope of \bhs, which in turn affects extrapolations of the relation to higher and lower mass systems.  To investigate the accuracy of current methods, black hole mass estimates for a number of objects, using multiple methods for each system, are required.
 
Several such tests have been conducted.  Cappellari et al. (2002) and Verdoes Kleijn et al. (2002) have constructed both gas dynamical models and axisymmetric stellar dynamical models for individual galaxies.  In both cases, the traditional thin-disc gas model requires a significantly (7 times and 30 times, respectively) smaller black hole than that needed in the stellar model, and the high stellar dynamical estimate is strongly excluded by the gas kinematics.  Also in both cases, the gas and stellar measurements bracket the black hole mass predicted by the \bhs\ relation.  To reconcile these results, potential sources of error in the gas and stellar models were examined, including non-gravitational motion in the nuclear gas discs, deviations from axisymmetry, and data quality.  The limitations of the data and of the assumptions in these studies require that further tests on a number of other galaxies be performed.

The nearby elliptical (E1) galaxy NGC~3379 (M105) is an ideal candidate for such a test.  NGC~3379 has been classified as a ``core" galaxy, in that the slope of its surface brightness profile is less positive in the central regions than in the rest of the galaxy (core radius of $R_c$=1\farcsec98, Lauer et al. 2005), and Faber et al. (1997) have suggested that such a surface brightness distribution can be generated through mergers that result in a central binary SMBH system.  Gebhardt et al. (2000b) have noted that the central two arcseconds (100 pc, at an assumed distance of 10.28 Mpc from Tonry et al. 2001) contain a well-defined dust disc, presumably linked to ionised gas and rotating around this SMBH (see also Pastoriza et al. 2000).  Additionally, extensive stellar kinematic data out to 90\arcsec (4.5 kpc) suggest that NGC~3379 has a very regular dynamical structure and may therefore be suitable for axisymmetric stellar dynamical models (Statler \& Smecker-Hane 1999; Gebhardt et al. 2000b; Emsellem et al. 2004).  From the stellar velocity dispersion measured in these data (201 \kms, Cappellari et al. 2006) and the \bhs\ equation given by Tremaine et al. (2002), NGC~3379 is expected to harbour a SMBH of mass $1.4 \times 10^8$ \msun; such a black hole has a sphere of influence of radius $R_{\rm BH} \approx\ \rbh\ $(15 pc).  It should therefore be possible, with data of sufficient spatial resolution, to construct both gas dynamical models and stellar dynamical models for this galaxy and generate two independent estimates of the mass of the central SMBH.

Because of the proximity, brightness, and apparent normalcy of NGC~3379, this galaxy has already been the subject of many studies.  While there are no previous gas dynamical models of the central regions of NGC~3379, Gebhardt et al. (2000b) have constructed axisymmetric stellar dynamical models, using a combination of ground-based long-slit data along four position angles and an FOS spectrum.  Their best-fitting model required a black hole of $1^{+1.0}_{-0.4} \times 10^8$ \msun\ (1$\sigma$ errors, Gebhardt et al. 2000b), consistent with the predictions of the \bhs\ relation.

In this paper, we study both the gas and stellar kinematics to generate two independent measurements of the mass of the SMBH of NGC~3379 and to investigate the structure of the stars and gas in its vicinity.  We use the {\tt SAURON} and {\tt OASIS} ground-based integral-field units (IFUs) to obtain stellar kinematics in two spatial dimensions (Section \ref{IFU}), and we use {\it Hubble Space Telescope} ({\it HST}) data to examine the surface brightness and kinematics of the central gas disc in detail (Section \ref{HST}).  In Sections \ref{StellarMod} and \ref{StellarResults}, the axisymmetric stellar dynamical model is presented, and the results are described.  Likewise, Sections \ref{GasMod} and \ref{GasResults} discuss the gas dynamical model and results, respectively.  We then compare these results (Section \ref{Discussion}) to gain quantitative insight into the robustness of black hole mass estimation techniques and summarise our conclusions in Section \ref{Conclu}.

\section{Observations: Ground-Based IFUs}
\label{IFU}

\subsection{{\tt SAURON} Spectroscopy}
\label{SAURON}

To study both the stellar and gas kinematics of NGC~3379, two complementary sets of integral-field data, from the {\tt SAURON} and {\tt OASIS} instruments, were obtained.  NGC~3379 is a part of the {\tt SAURON} survey of nearby early-type galaxies and bulges (de Zeeuw et al. 2002) and, as such, was observed out to $\sim 1 R_e$ using the {\tt SAURON} IFU (Bacon et al. 2001) at the 4.2-m William Herschel Telescope in February 1999.  The galaxy was observed in eight exposures, divided between three pointings, each including the galaxy centre.  Within pointings, individual exposures were spatially dithered in order to avoid systematic effects.  Characteristics of the {\tt SAURON} instrument and details of the observations are given in Table \ref{charac}.

The reduction process and reduced {\tt SAURON} datacube were presented in Emsellem et al. (2004).  Briefly, the data were reduced using the XSauron software and reduction process, developed at the CRAL-Observatoire and described in Bacon et al. (2001).  The steps included bias and dark subtraction, extraction of the spectra using a fitted mask model, wavelength calibration, low-frequency flat-fielding, cosmic-ray removal, homogenisation of the spectral resolution over the field, sky subtraction, and flux calibration.  All eight exposures were then merged into a single datacube and truncated to a common wavelength range.  During this process, the data were spatially resampled on to a common spatial scale of $0\farcsec8 \times 0\farcsec8$, resulting in a field-of-view of approximately $45\arcsec \times 70\arcsec$.  To increase the signal-to-noise (S/N) ratio to sufficient levels for accurate determination of the kinematics, the datacube was spatially binned using the Voronoi 2D-binning scheme developed by Cappellari \& Copin (2003).  A minimum S/N of 60 per spectral element was imposed; in practice, however, many of the spectra have a much higher S/N (S/N$_{\rm max} \approx$ 560), and over one quarter of the spectral elements remain unbinned.

Since we are interested in the details of the central regions of NGC~3379, a precise measurement of the PSF of the observations is necessary.  This was accomplished by comparing the reconstructed {\tt SAURON} intensity distribution to the {\it HST}/WFPC2/F555W data (see Section \ref{WFPC}), the filter that best matches the {\tt SAURON} spectral range.  The WFPC2 data were convolved with a PSF, which was modelled as the sum of two concentric circular Gaussians, and the difference between the resulting image and the {\tt SAURON} data was minimised.  The derived PSF parameters are listed in Table \ref{charac}.

\begin{table}
 \centering
 \begin{minipage}{140mm}
 \caption{Specifications of {\tt SAURON} and {\tt OASIS} instruments and observations.}
 \begin{tabular}{@{}lll@{}}
   \hline
   & {\tt SAURON} & {\tt OASIS} \\
  \hline
  Field of view & $33\arcsec \times 41\arcsec$ & $8\arcsec \times 10\arcsec$ \\
  Aperture size & 0\farcsec94 & 0\farcsec27 \\
  Final spatial sampling & 0\farcsec8 & 0\farcsec2 \\
  Spectral range & 4810 - 5300 \AA & 4760 - 5558 \AA \\
  Spectral sampling & 1.1 \AA/pixel & 1.95 \AA/pixel \\
  Spectral resolution (FWHM) & 4.2 \AA & 5.4 \AA \\ 
  Instrumental dispersion $\sigma$ & 108 \kms & 135 \kms \\
  Number of field lenses & 1431 & 1009 \\
  Number of sky lenses & 146  & 0 \\
  Number of pointings & 3 & 1 \\
  Number of exposures & 8 & 2 \\
  Time per exposure & 1800 s &  3600 s \\
  Weight, FHWM of PSF Gaussian 1 & 0.71, 1\farcsec60 & 0.84, 0\farcsec94 \\
  Weight, FHWM of PSF Gaussian 2 & 0.29, 5\farcsec22 & 0.16, 4\farcsec31 \\	
  \hline
 \label{charac}
 \end{tabular}
 \end{minipage}
\end{table}

\subsection{{\tt OASIS} Spectroscopy}
\label{OASIS}

NGC~3379 is also included in the {\tt OASIS} survey, which investigates the centres of the {\tt SAURON} galaxies with a similar spectral range to {\tt SAURON} but with higher spatial resolution (McDermid et al. 2004).  {\tt OASIS} integral-field observations of NGC~3379 were taken at the 3.6-m Canada-France-Hawaii Telescope in March 2001.  One pointing of two exposures was obtained, with the exposures dithered to provide oversampling and to avoid systematic effects.  The details of the {\tt OASIS} instrumental characteristics and of the observations are given in Table \ref{charac}.

The data were reduced using the publicly available XOasis software and reduction process, developed at the CRAL-Observatoire, described in Rousset (1992) and presented in McDermid et al. (2006).  The reduction process is nearly identical to that for the {\tt SAURON} data with two notable exceptions: the lack of sky subtraction (as {\tt OASIS} does not have dedicated sky lenslets) and the inclusion of a fringe correction.  Since {\tt OASIS} was pointed at the high-signal central regions of NGC~3379, the sky signal is so far below the galaxy signal as to be undetectable, and neglecting sky removal therefore has a negligible effect on the resulting reduced datacube.  Inspection of a preliminary reduction of the {\tt OASIS} data revealed that the EEV1 detector used in the observations suffered from the effects of fringing.  In order to remove this effect, an average twilight spectrum was divided off from a twilight datacube, leaving only the residual fringe pattern in all three dimensions (two spatial and one spectral).  The science datacubes were then divided by this fringe pattern.  The two final science datacubes were merged into a single datacube, during which the exposures were spatially resampled on to a common spatial scale of $0\farcsec2 \times 0\farcsec2$, resulting in a field-of-view of approximately $8\arcsec \times 10\arcsec$.  The data were spatially binned to a minimum S/N of 100 per spectral resolution element to accommodate the sparser spectral sampling of {\tt OASIS}.  In the central regions, this resulted in a maximum of two spatial elements being combined; since the PSF was oversampled, this mild binning did not cause a loss of spatial information.  The final {\tt OASIS} datacube of NGC~3379 is presented in McDermid et al. (2006).

The PSF of the {\tt OASIS} measurements of NGC~3379 was determined using the same procedure as used for the {\tt SAURON} data, and the derived PSF parameters are listed in Table \ref{charac}.

\subsection{Stellar Kinematics}
\label{StellarKin}

The {\tt SAURON} and {\tt OASIS} datacubes contain several absorption features, including \hb, \mgb, and some Fe lines, as well as several emission lines, including \hb, \oiii~$\lambda \lambda$4959,5007, and \nidoub~$\lambda$5200.  To extract reliable stellar kinematics, we used the penalised pixel fitting (pPXF) method of Cappellari~\&~Emsellem~(2004), which allows the emission regions to be omitted from the fit and minimises the effects of undersampling via penalisation.  An optimal template is constructed from a stellar library and is then convolved with Gauss-Hermite expansions (van der Marel \& Franx 1993; Gerhard 1993) out to six terms until differences between the convolved template and the observed galaxy spectra are minimised.

To estimate the error on the stellar kinematics, Monte Carlo simulations of the kinematic extraction were performed, following the prescriptions of Cappellari \& Emsellem (2004).  Spectra were simulated by adding noise, based on the residuals of the original extraction, to the galaxy spectra.  During the Monte Carlo realisations, the pixel fitting was done without penalisation, in order to obtain the unbiased scatter of the kinematics.  One hundred realisations were used, and the 1$\sigma$ errors on the kinematics were derived from the resulting distributions.

This method is similar to that used by Emsellem et al. (2004) on the {\tt SAURON} datacube and by McDermid et al. (2006) on the {\tt OASIS} datacube, but it differs in the details of the extraction.  In those papers, optimal stellar templates are constructed from a selection of single-burst stellar population models by Vazdekis (1999) and additional strong \mgb\ stellar spectra from the Jones (1997) library.  The Jones spectra were used to attempt to compensate for possible \mgb\ over-abundance and thus to alleviate potential template mismatch.  Both Emsellem et al. (2004) and McDermid et al. (2006) then simultaneously fit the optimal template and the Gauss-Hermite moments out to four terms for each bin.

Preliminary tests revealed that the possibility for template mismatch could be further reduced through the use of the larger MILES (Mid-resolution INT Library of Empirical Spectra) library of 998 individual stellar spectra, kindly provided pre-publication by S\'anchez-Bl\'azquez et al. (2006).  This library has several advantages over the combined Vazdekis and Jones library, the most notable being greater coverage and sampling of stellar atmospheric parameters such as metallicity.  This variety of abundances and larger number of stars in the library allows for a more precise determination of the optimal template, which for NGC~3379 is composed of 16 of the 998 spectra.  We quantify the improvement in the optimal template through the RMS scatter of the residuals from a pPXF fit to a high S/N {\tt SAURON} spectrum of this galaxy; the use of the MILES library instead of the Vazdekis and Jones library reduces this scatter from $\sim 0.55$ to $\sim 0.20$.  Emsellem et al. (2004) show that the maximum systematic error in the higher Gauss-Hermite moments from template mismatch scales almost linearly with the RMS scatter (see their Fig. B3).  This implies that the MILES library decreases the maximum systematic errors by a factor of $\sim 3$, to less than 0.02.

\begin{figure*}
	\includegraphics[width=17cm]{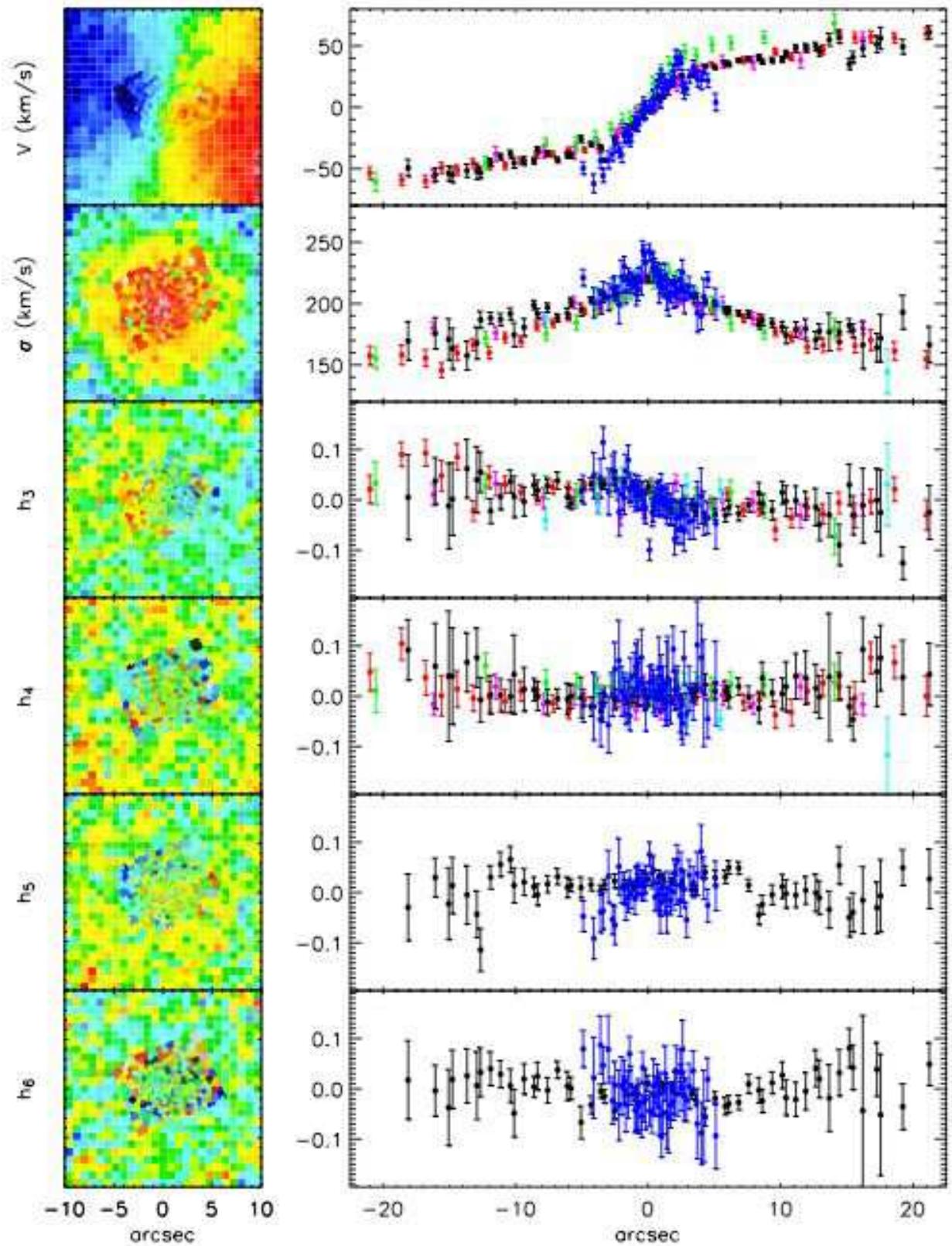}
	\caption{First six Gauss-Hermite moments of the LOSVD in NGC~3379.  Left panels show the inner ten square arcseconds of the {\tt SAURON} field, with the {\tt OASIS} field overlaid.  In these plots, north is up and east is to the left.  Right panels show the LOSVD moments as seen through a 1\arcsec\  slit placed along the galaxy's major axis (PA=70$^\circ$) as observed with {\tt SAURON} (black) and {\tt OASIS} (blue).  Red, magenta, cyan, and green points indicate the long-slit major axis data of Statler \& Smecker-Hane (1999), Gebhardt et al. (2000b), Halliday et al. (2001), and Samurovi\'{c} \& Danziger (2005) respectively. The small {\tt OASIS} PSF allows the data to probe the central rise in velocity dispersion with more detail than in previous measurements or in the {\tt SAURON} data.}
	\label{SAUandOASkin}
\end{figure*}

Using the MILES optimal template, the kinematics in each bin were extracted out to six Gauss-Hermite terms for both the {\tt SAURON} and {\tt OASIS} data.  For maximum consistency between these two data sets, the {\tt OASIS} spectra were truncated to the same spectral region as the {\tt SAURON} spectra (roughly from \hb\ to \mgb) for the kinematic extraction.  This was desirable because of the number of additional Fe features included in the {\tt OASIS} spectral range, which render those data particularly sensitive to potential inabilities of the template library to reproduce the exact abundance ratios of this galaxy (see also Barth, Ho \& Sargent 2002).  The results of this extraction are consistent with the kinematics presented in Emsellem et al. (2004) and McDermid et al. (2006), with the exception that the measurements of the even moments presented here are slighly lower, a testament to the decreased effect of template mismatch through the use of the more extensive stellar library.

A comparison of the {\tt SAURON} and {\tt OASIS} kinematics obtained with this method revealed nearly perfect agreement between all moments except the velocity dispersion, in which the {\tt OASIS} data were  approximately 5 \kms\ higher than the {\tt SAURON} data.  Given the excellent consistency in the higher moments, this deviation is most likely not an effect of template mismatch but rather of a systematic error in the reduction process.  The most probable source of error is the homogenisation of the spectral resolution across the field, a step that involves convolution with a Gaussian.  To correct for this offset, we measure the circularly averaged velocity dispersion profiles between 3\arcsec and 4\arcsec, outside the influence of the PSFs.  Since the systematic uncertainties in the homogenisation are similar for both data sets, we adjust both sets to the midpoint of these radial profiles (in practice, $\pm$2.5 \kms) and implement the correction in quadrature to the entire dispersion field.  This correction, used only for the velocity dispersion, mimics the nature of the problem as an error in the Gaussian convolution during the resolution homogenisation.

The resulting {\tt SAURON} and {\tt OASIS} kinematics are shown and compared in Fig. \ref{SAUandOASkin}.  (The full {\tt SAURON} field is shown in Section \ref{StellarResults}.)  Long-slit stellar kinematics for this galaxy have been measured by Statler \& Smecker-Hane (1999), Gebhardt et al. (2000b), Halliday et al. (2001), and Samurovi\'{c} \& Danziger (2005), all of whom extracted their LOSVD out to $h_4$.  Figure \ref{SAUandOASkin} also presents the comparison of the {\tt SAURON} and {\tt OASIS} major axis kinematics to those long-slit measurements along the same axis.  Overall, the kinematics agree quite well with each other and with the literature.

\begin{figure}
	\includegraphics[width=8cm]{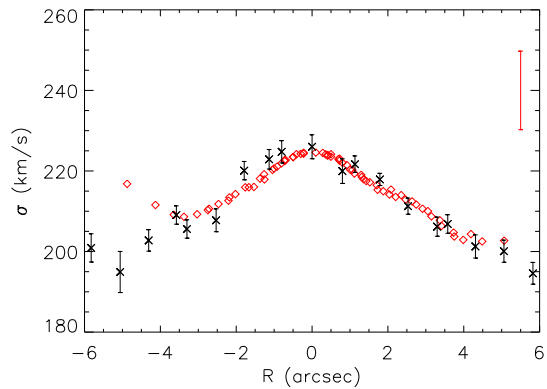}
	\caption{{\tt OASIS} stellar velocity dispersions (diamonds) are convolved up to the PSF of the {\tt SAURON} observations and compared to the {\tt SAURON} dispersion measurements (crosses), both shown as extracted from slits along the major axis.  The bar in the upper right corner gives the mean error of the unsmoothed {\tt OASIS} data.  The convolved {\tt OASIS} values are given for all the original bins and therefore show strong point-to-point correlations.  In general, this convolution brings the two data sets into agreement.}
	\label{conv}
\end{figure}

This qualitative agreement is reassuring; however, these kinematics will be used in detailed stellar dynamical modelling, which requires that they are completely consistent and are free of systematic effects.  We therefore must establish that the somewhat ad hoc correction to the velocity dispersion measurements (as determined using data between 3\arcsec and 4\arcsec) produces consistent velocity dispersions throughout the field.  At small radii, however, the {\tt OASIS} dispersions are expected to be noticeably higher, since the narrower {\tt OASIS} PSF allows the data to better resolve the rising central velocity dispersion (Fig. \ref{SAUandOASkin}).  Since we have detailed measurements of the {\tt SAURON} and {\tt OASIS} PSFs, we can test the effectiveness of the dispersion correction by convolving the {\tt OASIS} velocity dispersions up to the {\tt SAURON} PSF and then comparing them to the observed {\tt SAURON} velocity dispersions.  To accomplish this, we convolve the flux-weighted true moments ($\mu_1$ = $v$ and $\mu_2^2$ = $v^2 + \sigma^2$) and then subtract the convolved true moments in quadrature to find the convolved dispersion (see Emsellem, Monnet \& Bacon 1994 for a more complete discussion of this method).  The results of this comparison are shown in Fig. \ref{conv} and indicate that the two data sets are in good agreement.  Edge effects from the convolution are visible as slight rises in velocity dispersion at the boundaries of the {\tt OASIS} field, but most important for the determination of the SMBH mass, the central dispersion values are consistent.  This agreement, as well as the agreement with the literature shown in Fig. \ref{SAUandOASkin}, suggests that the steps taken here to mitigate potential effects of template mismatch have been largely successful and have produced kinematics that adequately describe this galaxy.

\subsection{Gas Kinematics}
\label{IFUGasKin}

In the region of the nuclear gas disc, the {\tt SAURON} and {\tt OASIS} datacubes show strong emission features, which can be used to measure the ionised gas kinematics in the galaxy centre.  In the {\tt SAURON} field, however, the 0\farcsec8 pixels probe the $\sim$ 2\arcsec-diameter disc only marginally.  Sarzi et al. (2006) present the ionised gas kinematics derived from the {\tt SAURON} datacube, in which the disc's rotation is visible in only the central few pixels.  We therefore make use of the {\tt OASIS} datacube, with 0\farcsec2 pixels, to obtain well-sampled spatially complete gas kinematics of this disc.

\begin{figure*}
	\includegraphics[width=17cm]{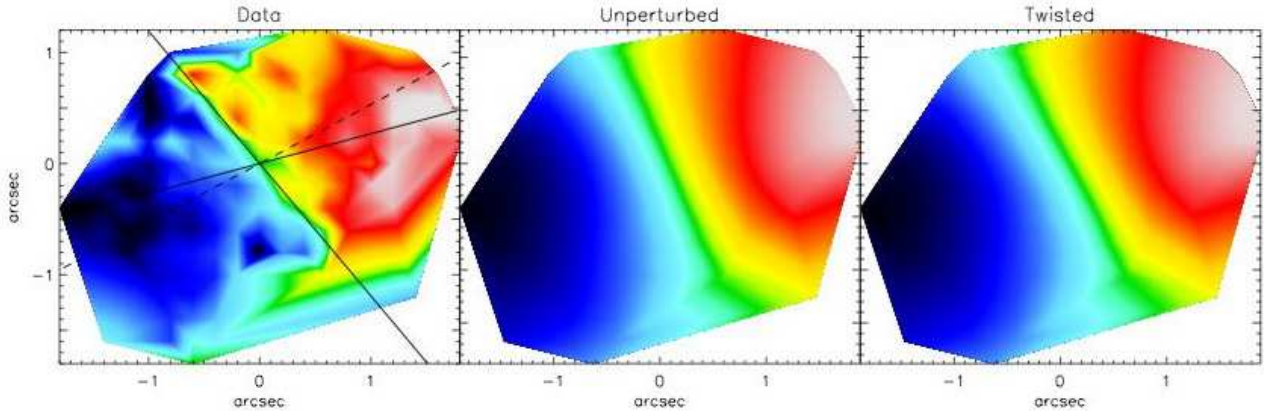}
	\caption{{\tt OASIS} observed gas velocities (left) with the kinematic major and minor axes traced by solid black lines and the photometric major axis indicated with the dashed line (north up, east to the left).  The unperturbed model (centre) was constructed with the data symmetrised around the kinematic major axis, since a determination of M/L depends strongly on the peak velocities along the major axis.  The twisted model (right) is only a marginally better fit to the data, since most of the features of the twist are smeared out by the {\tt OASIS} PSF.  However, qualitative analysis of the data strongly favours some type of perturbed model, as the kinematic major and minor axes are not orthogonal. The data shown in the left panel have been interpolated between pixels to clarify the kinematic axes and for ease of comparison with the model; the original kinematics are presented without interpolation in McDermid et al. (2006).}
	\label{BarthOasis}
\end{figure*}

The gas kinematics in the {\tt OASIS} datacube are measured following the method of Sarzi et al. (2006), which McDermid et al. (2006) have implemented on the {\tt OASIS} sample.  In this method, the pixel fitting is rerun with the stellar kinematics held fixed at the values determined above.  Rather than excluding spectral regions around the emission lines, however, the emission lines are included in the template library as a series of Gaussians (one per line).  The velocity and broadening of the emission lines are determined entirely from the \oiii\  emission and then imposed on the other emission features.  The amplitudes of all lines are then derived via an optimal (non-negative) linear combination of stellar spectra and emission line templates.

To remove spurious detections, we conservatively require an amplitude-to-noise ratio of 5 before an emission line is considered real (see McDermid et al. 2006, for a full discussion of sensitivity limits). The resulting gas kinematics of sufficient significance are shown in Fig. \ref{BarthOasis}.  In these data, the shape and orientation of the nuclear gas disc are visible, as is the lack of ionised gas outside this disc.

The {\tt OASIS} data are limited, however, by the seeing, which smears emission from the central 2\arcsec of the galaxy over $\sim$ 4\arcsec.  Thus, although the spatial resolution of the {\tt OASIS} instrument is sufficient to probe the sphere of influence of a hypothesised central black hole, the PSF smoothes these effects, such that any Keplerian rise in velocities is undetectable.  While extremely useful for studying the complete velocity field of the gas disc, the {\tt OASIS} data alone are insufficient for a gas dynamical model to measure the black hole mass in this galaxy.

\section{Observations: {\it HST}}
\label{HST}

\subsection{WFPC2 Imaging}
\label{WFPC}

To further investigate the central gas disc in NGC~3379, spatially-resolved photometric and spectroscopic data were obtained with {\it HST}, program 8589.  The central region of the galaxy was imaged using the Wide Field Planetary Camera 2 (WFPC2), with the galactic nucleus placed on the higher-resolution PC chip (0\farcsec0455/pixel).  Data were obtained through the F658N filter in order to observe the \ha\  and \nii\  emission in the central gas disc.  With a spectral range from $\sim$ 6575-6605 \AA, this passband includes all the \ha\  emission and most of the \nii\  emission at this redshift ($\sim$ 900 \kms).

The F658N image was taken as a sequence of two individual exposures, each of length 1200 seconds.  The images were processed in the {\it HST} data pipeline and were subsequently combined and cosmic-ray rejected.

\begin{figure}
	\includegraphics*[width=8.5cm]{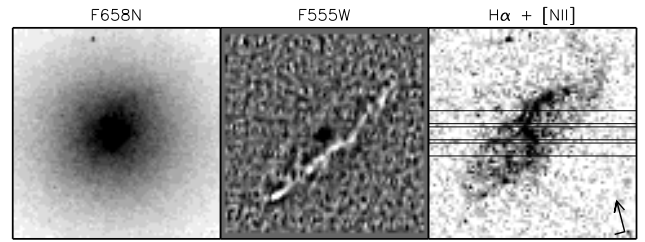}
	\caption{{\it HST}/WFPC2 images of the centre of NGC 3379.  The \ha\  narrow band continuum emission is shown at left, the dust ring (centre) is detected in an unsharp-masked {\it V}-band image, and the \han\ image (right) was generated by the subtraction of the \ha\ image by an ``{\it R}"-band continuum image (see text).  The nominal location of the three STIS slits is shown in the right panel, with the position angle of the images indicated by the arrow on the lower right.}
	\label{WFPCbands}
\end{figure}

In addition to these data, Gebhardt et al. (2000b) obtained F555W ({\it V}-band) and F814W ({\it I}-band) images of the nucleus of NGC~3379.  In both of these images, the galaxy was also centred on the PC detector.  Gebhardt et al. (2000b) present the deconvolved F555W image, with the galaxy's spheroidal light distribution subtracted, to reveal a central dust disc seen in absorption (see also Lauer et al. 2005).  This disc corresponds in orientation and inclination with the gas emission in \ha, as shown in Fig. \ref{WFPCbands}.

To generate an image of the gas disc without stellar emission, we scale a broadband image and subtract it from the F658N data.  In the absence of any {\it HST} {\it R}-band data, we simulate data from this passband by scaling the F555W and F814W images and averaging the two.  This method yields a more accurate representation of the continuum emission and dust absorption present in the F658N image than can be achieved with either the {\it V}-band or {\it I}-band image alone.  The resulting ``{\it R}-band" image was subtracted from the F658N data to produce a map of only the \han\ emission (Fig. \ref{WFPCbands}). 

The resulting \han\  image reveals a well-defined disc of emission around the galaxy centre.  The regularity of this disc allows for a simple and precise measurement of its photometric position angle and inclination.  We fit an ellipse by eye to the outline of the disc in this image and measured the position angle to be PA\ =\ 118$^\circ$.  If the disc is assumed to be thin and intrinsically circular, then the inclination implied by this fit is $i_{\rm disc} = 70^\circ \pm 8^\circ$, where the error bar includes the inclination of the dust disc associated with the gas and an ellipse fit to the {\tt OASIS} gas data.

\begin{figure}
\begin{center}
	\includegraphics[height=5cm]{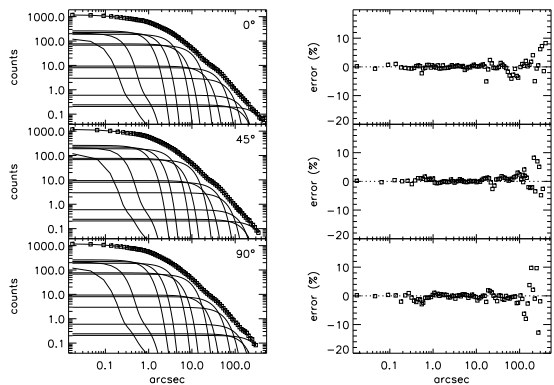}
	\caption{MGE fit to the data along three position angles (left), with percentage error (right).  All individual Gaussian components are shown, as well their sum, which fits the surface brightness data (boxes) over four orders of magnitude in radius.  This model was generated from combined {\it HST}/WFPC2 data and ground-based wide-field MDM imaging and was presented in Cappellari et al. (2006).  The strong ``core" signature can be seen inwards of $\sim$ 2\arcsec.}
	\label{MGEmodel}
\end{center}
\end{figure}

Cappellari et al. (2006) have used the F814W WFPC2 image of NGC~3379, in addition to wide-field ground-based photometry taken in the same filter at the 1.3-m McGraw-Hill telescope at the MDM observatory on Kitt Peak, to construct a Multi-Gaussian Expansion (MGE) parametrization of the surface brightness of this galaxy (Emsellem et al. 1994; Cappellari 2002).  Their deconvolved MGE model was regularised to require that the axial ratio of the flattest Gaussian be as large as possible and was corrected for extinction.  The resulting best-fitting MGE, a sum of 13 Gaussian components, is presented in their Fig. 3, and the calibrated parameters, corrected for extinction and converted to a stellar surface density in solar units, are given in their Table B1.  In Fig. \ref{MGEmodel}, we show the MGE fit to the surface brightness along several position angles, as well as the residuals of this fit.  We adopt this parametrization to describe the stellar surface density distribution in this galaxy.

\subsection{STIS Spectroscopy}
\label{STIS}

Long-slit spectra of the central gas disc in NGC~3379 were obtained with the {\it HST} Space Telescope Imaging Spectrograph (STIS), through a 0\farcsec2 slit and through the G750M filter with the grating tilted to provide a spectral range 6300 - 6860 \AA.  With this set-up, these data included the H$\alpha$ and both [NII] emission lines in this galaxy, at a spectral resolution of 0.554\ \AA/pixel and a spatial sampling of 0\farcsec051/pixel.

Target acquisition and peak-up procedures were performed to centre the STIS slit on the galaxy's nucleus.  Spectroscopic images were then obtained at each of three locations: at the galaxy centre and with a nominal offset $\pm0\farcsec25$ perpendicular to the slit, with a resulting gap of 0\farcsec05 between slits.  Due to the scheduling of the observations, there was no observability at the requested position angle along the major axis of the gas disc; consequently, the observations were taken at a PA = -104.7$^\circ$, approximately 40$^\circ$ offset from the major axis.  This slit positioning and location is shown in Fig. \ref{WFPCbands}, and we designate the ``top," ``centre," and ``bottom" slits according to their location in this figure.

\begin{figure*}
	\includegraphics[width=15cm]{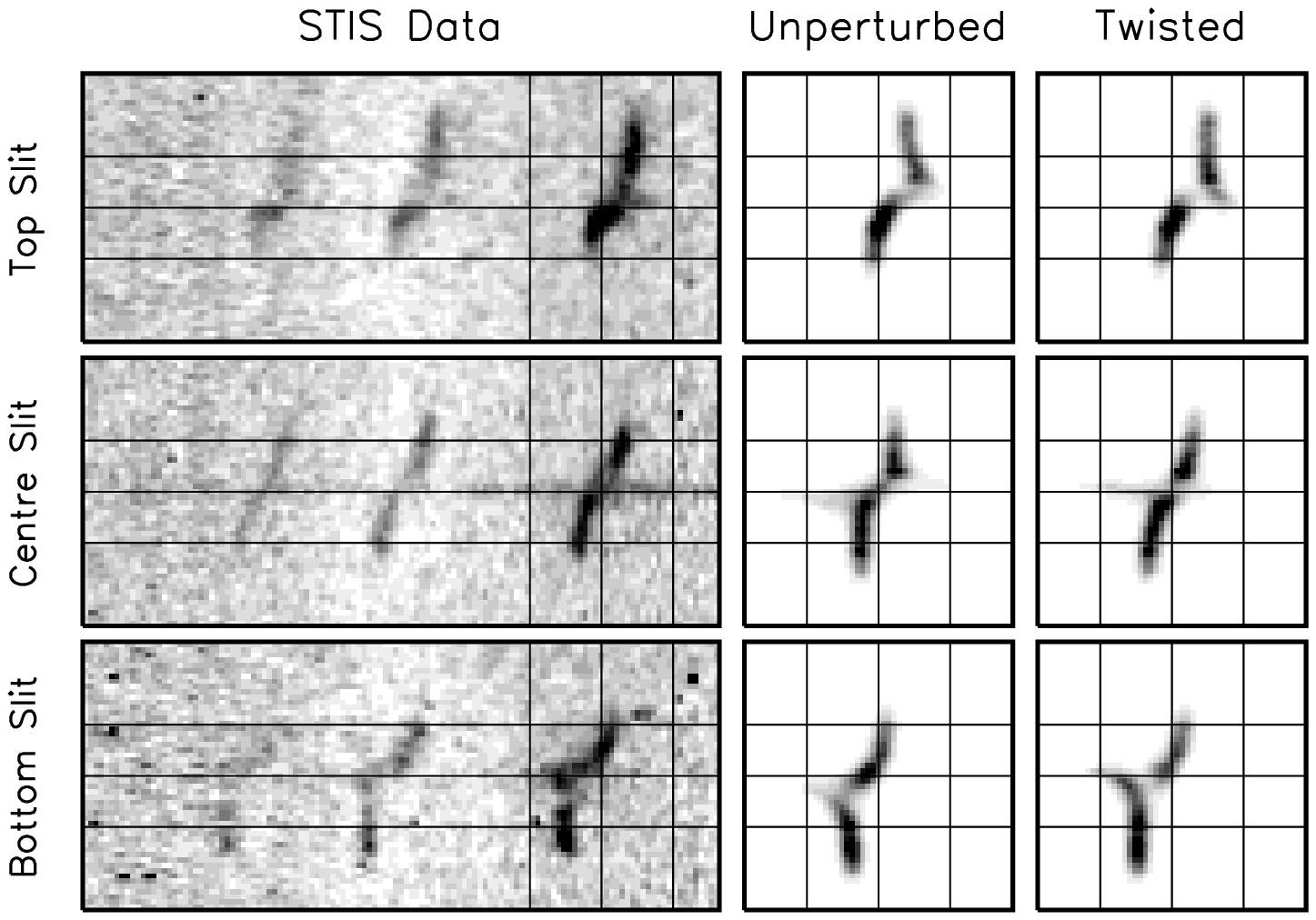}
	\caption{STIS longslit observations for all three slits (left), unperturbed model (centre), and a twisted model (right).  Lines have been added to guide the eye.  In all images, wavelength is on the horizontal axis and distance on the vertical axis, with the galaxy centre along the middle horizontal line.  In the STIS data, the emission lines from left to right are \nii\ $\lambda 6548$, \ha, and \nii\ $\lambda 6584$.  The main differences between the two models are the spatial positions of the peak velocity in the top and bottom slits, the shape of the spatial variation in velocity, and the ability to reproduce intensity variations across the slits.  See text for further discussion of the models.}
	\label{BarthFig4}

	\includegraphics[width=15cm]{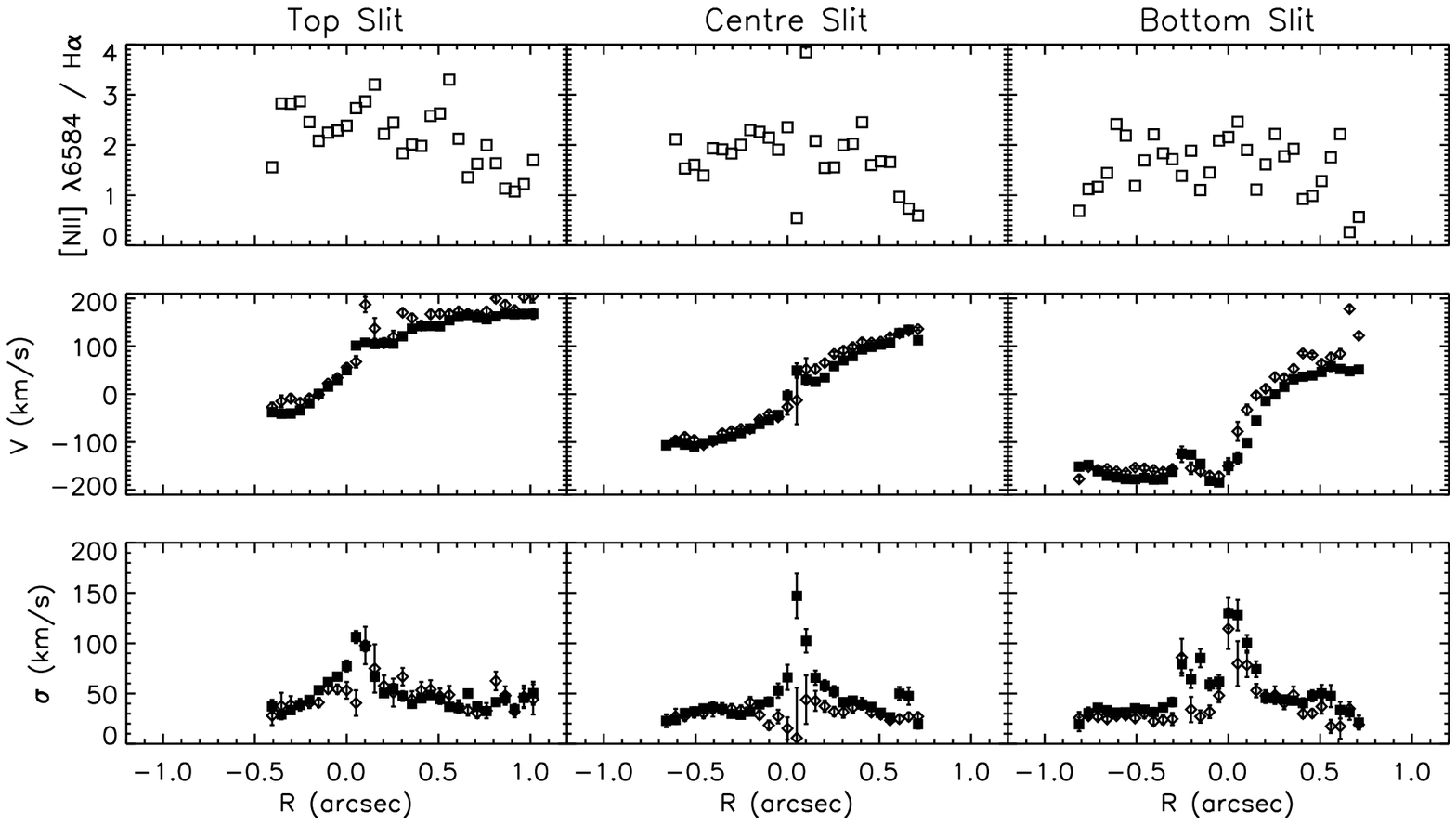}
	\caption{Results of kinematic extraction for the \nii\ $\lambda$6584 (closed squares) and \ha\  (open diamonds) emission lines, for all three slits.  Shown here are the ratio of \nii\ $\lambda$6584 emission line flux to \ha\  emission line flux (top row), and the extracted velocities (middle row) and velocity dispersions (bottom row).}
	\label{STISkin}
\end{figure*}

Data were acquired in five exposures on the top and centre slits and four on the bottom slit.  Observation details are presented in Table \ref{STISobs}.  Individual exposures were spatially dithered along the slit in order to avoid systematic effects.  The images were processed by the {\it HST} data processing pipeline, including wavelength calibration, after which they were combined and cosmic-ray rejected.  The resulting STIS data for all three slits is shown in Fig. \ref{BarthFig4}.

We use the STIS data themselves to measure the actual positions of the slits by collapsing the spectra over the spectral range and comparing that light profile to light profiles extracted from the acquisition image taken during the peak-up procedure.  These latter profiles were extracted from the acquisition image by averaging the flux of pixels (and fractions of pixels) covered by synthetic 0\farcsec2 STIS slits.  The best-fitting slit positions were determined using a $\chi^2$ minimization of the ratio between the light profile measured from the slits and that measured from the acquisition image.  For the top and centre slit, the nominal offset is indeed the actual offset, to within a fraction of a STIS pixel.  For the bottom slit, the actual offset is -0\farcsec20, which deviates from the nominal offset of this slit by $\sim$1 STIS pixel.  These results are summarised in Table\ \ref{STISobs}.

\begin{table}
 \centering
 \begin{minipage}{140mm}
 \caption{Specifications of the {\it HST}/STIS observations.}
 \begin{tabular}{lcccc}
  \hline
    & \# Exp. & Avg. $t_{\rm Exp}$ & Nominal Shift & Actual Shift \\
  \hline
  Top Slit & 5 & 2600 s & +0\farcsec25 & +0\farcsec25 \\
  Centre Slit & 5 & 2700 s & 0\farcsec00 & 0\farcsec00 \\
  Bottom Slit & 4 & 2550 s & -0\farcsec25 & -0\farcsec20 \\
  \hline
 \label{STISobs}
 \end{tabular}
 \end{minipage}
\end{table}

\subsection{Gas Kinematics}
\label{GasKin}

The gas kinematics in NGC~3379 were extracted for each individual row of all three STIS slits.  As would be expected from the WFPC2 \han\  image, the emission signal drops off dramatically at the boundaries of the gas disc; therefore, all rows with detectable signal were used.  This results in gas kinematics being obtained for roughly the inner two arcseconds of the galaxy.

To measure the kinematics, we first removed the stellar background emission by fitting a linear continuum to 6475 - 6525 \AA\ and 6625 - 6675 \AA.  Over this spectral range, a straight line is a good approximation to the continuum shape.  The \ha\  and the \nii\ $\lambda$6584 lines were then fit separately by single Gaussians.  The \nii\ $\lambda$6548 emission was not used, as it has low signal in all rows and is not detectable at larger radii (see Fig. \ref{BarthFig4}).  No other emission lines were detected.  The higher \nii\ $\lambda$6584 signal than \ha\  and the lack of other significant emission lines are consistent with the classification of NGC~3379 as a LINER galaxy.

Figure \ref{STISkin} shows the derived \nii\ $\lambda$6584 and \ha\  kinematics, set to a systemic velocity of zero.  Over most of the disc, the two lines share roughly the same velocity distribution, although for over half the data points, their differences are higher than the formal errors.  The \ha\  velocity dispersion, in particular, is consistently lower than that of \nii.  A kinematic extraction that links the velocity and/or dispersion of the \nii\  component of this disc to that of \ha\  component would therefore introduce a bias in the kinematics.  We consequently restrict our kinematic measurements to those derived from the highest signal line, that of \nii\ $\lambda$6584.

\subsection{Stellar Kinematics}
\label{FOSandSTIS}

Using the pPXF method, we attempted to measure the central stellar kinematics using the STIS data, after careful masking of the emission features.  However, the kinematic extraction has to rely mainly on the \ha\ absorption feature, which is almost completely filled with gas emission.  For this reason, the accuracy on these measurements is modest and very sensitive to systematic effects, making the error bars significant ($\pm$40 \kms\ in the velocity dispersion).

Gebhardt et al. (2000b) also obtained {\it HST} stellar kinematics in the centre of NGC~3379 with an FOS pointing, from which they measured $v = 0 \pm 14$ \kms, $\sigma = 289 \pm 13$ \kms, $h_3 = 0.11 \pm 0.04$, and $h_4 = 0.04 \pm 0.04$.  They note the skewness of the LOSVD and attribute the non-zero $h_3$ value to the combined effects of miscentring of the aperture during the peak-up procedure and significant dust absorption in the central regions of the galaxy.  Interpreting these kinematics is therefore a somewhat complicated process.

The STIS and FOS stellar kinematics are not inconsistent with the {\tt SAURON} and {\tt OASIS} measurements described in Section \ref{StellarKin}; however, the large errors in the STIS kinematics and the effects of dust in the FOS kinematics limit their usefulness in stellar dynamical modelling.  Thus while the STIS emission line kinematics are critical for our gas dynamical modelling in Section \ref{GasMod}, the absorption line kinematics add few constraints, and much uncertainty, to the stellar models, and we do not include them in those models.

\section{Stellar Dynamical Modelling: Method}
\label{StellarMod}

\subsection{Axisymmetric Model Construction}
\label{Schw}

To model the orbital structure of NGC~3379 and obtain a stellar dynamical black hole mass estimate, we construct general stationary axisymmetric models of the stellar component of the galaxy.  These three-integral models are based on the Schwarzschild (1979, 1982) orbital superposition method, further developed by Richstone \& Tremaine (1988), Rix et al. (1997), and van der Marel et al. (1998).  This technique has been widely used to generate axisymmetric models of galaxies to determine galaxies' SMBH mass, mass-to-light (M/L) ratio, and orbital distribution and is thoroughly described in the literature by the aforementioned authors as well as by other groups (e.g., Gebhardt et al. 2003; Thomas et al. 2004; Valluri, Merritt \& Emsellem 2004; Cappellari et al. 2006). 

We use the implementation of Cappellari et al. (2006), with which they modelled a subsample of the {\tt SAURON} early-type galaxies.  This method proceeds in four steps.  First, the MGE parametrization is analytically deprojected and converted to a stellar potential, using an assumed shape (axisymmetry) and a constant stellar M/L ratio.  In the second step, a representative orbit library is constructed by integrating a large number of orbits in this potential.  Orbits are sampled evenly across observable space, such that their starting points are also varied over the three integrals of motion (energy $E$, angular momentum $L_z$, and non-classic third integral $I_3$).  This grid covers $>$99\% of the total luminous mass of the galaxy, with the radii sampled logarithmically from 0.01\arcsec to 400\arcsec.  During the third step, these orbits are integrated and projected into the plane of observables, namely the $x'$ and $y'$ locations on the sky and the line-of-sight velocity distribution.  This is done taking into account convolution with the PSF and aperture binning.  The final step consists of using non-negative least-squares (NNLS, Lawson \& Hanson 1974) fitting to determine the set of weights for each orbit such that the sum of the orbits in a given bin best reproduce the observed kinematics and stellar density in that bin.

This model implementation has been tested by Krajnovi\'{c} et al. (2005), who used it on {\tt SAURON} observations to measure the inclination, M/L ratio, and integral space structure of NGC~2974.  They evaluated the robustness of this method in recovering input parameters with a constructed two-integral ($E$ and $L_z$) test galaxy.  While they found that the model effectively recovers mass-to-light ratio and orbital structure, they note that the determination of their test galaxy's inclination is potentially degenerate, in that the differences between models of various inclinations are small with respect to realistic observational error estimates.

\subsection{Application to NGC 3379}
\label{Schw3379}

In the case of NGC~3379, the {\tt SAURON} data consist of 1603 bins and the {\tt OASIS} data of 344 bins.  Prior to modelling, anomalous bins are flagged to be omitted from symmetrisation and fitting.  In practice, two {\tt SAURON} bins (with a foreground star) and one {\tt OASIS} bin (with a bad pixel in the middle of the Mg {\it b} feature) are omitted.  With 6 LOSVD moments and the intrinsic and projected mass at each bin, there are thus 15552 observables.  These data are then symmetrised using the mirror-(anti-)symmetry of the kinematic field and averaging the kinematic values from four symmetric points around the major and minor axes.  Due to the irregular shape and distribution of the Voronoi bins, there are sometimes no bins at one of the symmetric positions; in this case, the value of the position is interpolated from surrounding bins.  During this process, the error bars remain unchanged, since the number of data points has not been altered.  The full, symmetrised {\tt SAURON} and {\tt OASIS} fields are presented with the model results in the following section.

As in Cappellari et al. (2006), the Schwarzschild model uses an orbit library constructed with 21 energies, and 8 angular and 7 radial values at each energy.  We employ a `dithering' method, which combines a large number of orbits (6$^3$ = 216) with different but adjacent initial conditions to allow the model to reproduce the observed (smooth) distributions.  The final galaxy model is thus made with 21 x (8-1) x 7 x 2 x 6$^3$ = 444,528 different orbits, divided into 2058 groups of 216 orbits before the NNLS fitting. (The (8-1) accounts for duplication of angular values when symmetries are considered, and the additional factor of 2 accounts for positive and negative $L_z$.)  We follow Verolme et al. (2002) in adopting a regularisation parameter of $\Delta=4$, which imposes smooth variation in integral space.  Given recent concerns about the effects of regularisation (e.g. Valluri et al. 2004; see Section \ref{StellarIssues} for a full discussion), we also generate a set of test models with no regularisation ($\Delta=1\times10^3$) to ensure that the derived black hole mass is not dependent on this parameter.

Each dynamical model requires three input parameters, namely the galaxy's inclination, the stellar M/L ratio, and the mass of the central SMBH.  The values of these parameters are estimated by constructing models with various values for all three parameters and then determining which model best reproduces the stellar kinematics and mass distribution in the $\chi^2$ sense.  In the case of NGC~3379, several stellar dynamical models have already been constructed, some of which indicate that this galaxy is likely seen edge-on (Gebhardt et al. 2000b).  However, the inclination of this galaxy has been the source of much debate, including suggestions that it is instead a face-on S0 (Cappaccioli et al. 1991; Statler \& Smecker-Hane 1999; see also Section \ref{StellarIssues}).  Subsequent models by Statler (2001), on the other hand, find that the most probable inclination is more moderate, at $i$$\approx$40$^\circ$.  With the MGE surface brightness parametrization, we can set a lower limit on the inclination of $i$=35$^\circ$ from the deprojection of the narrowest Gaussian component.  Following previous stellar dynamical results, we construct our primary set of models at $i$=88$^\circ$ (to avoid possible numerical artefacts at exactly 90$^\circ$), but we supplement these with relatively face-on models at $i$=50$^\circ$ to test the effects of inclination on the black hole mass measurement.  In both sets of models, we vary the M/L ratio and \mbh\ to fit the combined {\tt SAURON} and {\tt OASIS} data sets and investigate the internal structure of NGC~3379.

\begin{figure*}
	\includegraphics[width=17cm]{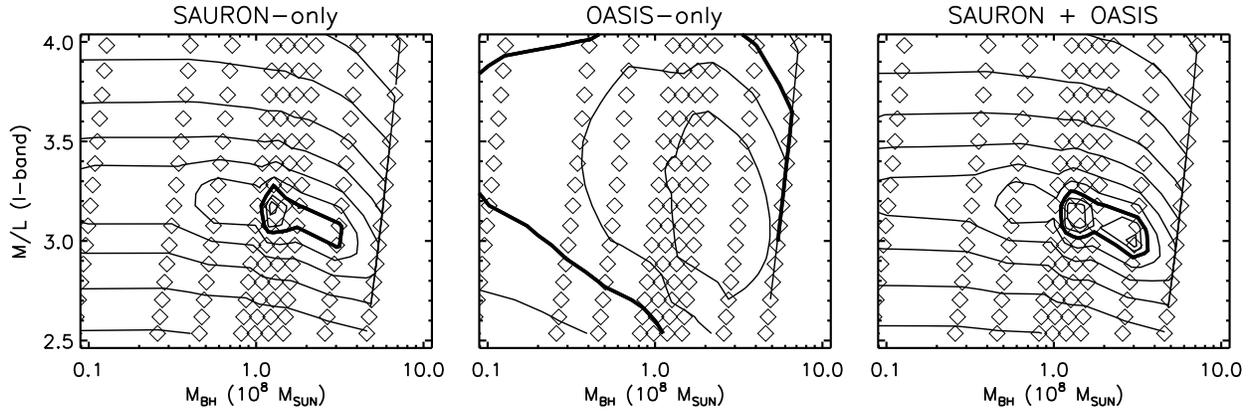}
	\caption{$\chi^2$ contours for the {\tt SAURON}-only models (left), {\tt OASIS}-only models (middle), and the models using the full, combined data set (right).  M/L is measured in solar units.  Diamonds indicate locations where models were constructed.  The thin lines are contours of integer values of the standard deviation, and thick contours indicate 3$\sigma$.  Models at \mbh=0 were also constructed but are not shown here; in all cases, those models are excluded to at least 4$\sigma$.}
	\label{SAUandOASchi}
\end{figure*}

\section{Stellar Dynamical Modelling: Results}
\label{StellarResults}

\subsection{Best-fitting Models}
\label{StarsBest}

\begin{figure*}
	\includegraphics[width=17cm]{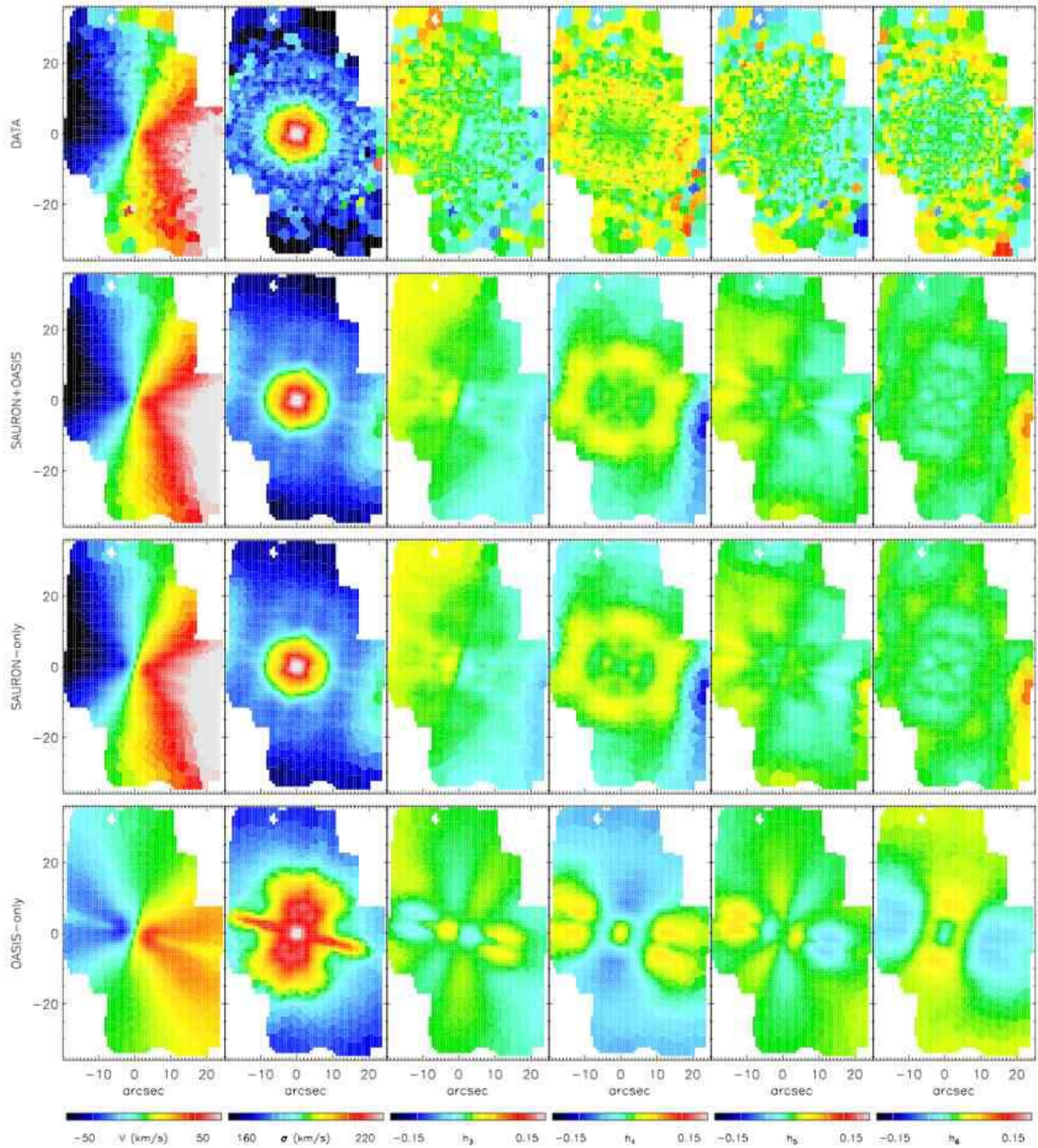}
	\caption{From top to bottom: The {\tt SAURON} symmetrised kinematic data, the best-fitting model with both the {\tt SAURON} and {\tt OASIS} data fitted, the model that best fits only the {\tt SAURON} data, and the predictions for the {\tt SAURON} field from the model that best fits only the {\tt OASIS} data.  The model that uses only the {\tt SAURON} data differs little from the model using the combined data set.  The model that uses the {\tt OASIS} data set suffers greatly from the lack of constraints outside 5\arcsec and consequently generates an unphysical orbital structure.  (Note that the high velocity region at $x \sim 0\arcsec$, $y \sim -20\arcsec$ in the data is a foreground star and is masked during the modelling.)}
	\label{SAUbest}
\end{figure*}

 \begin{figure*}
	\includegraphics[width=17cm]{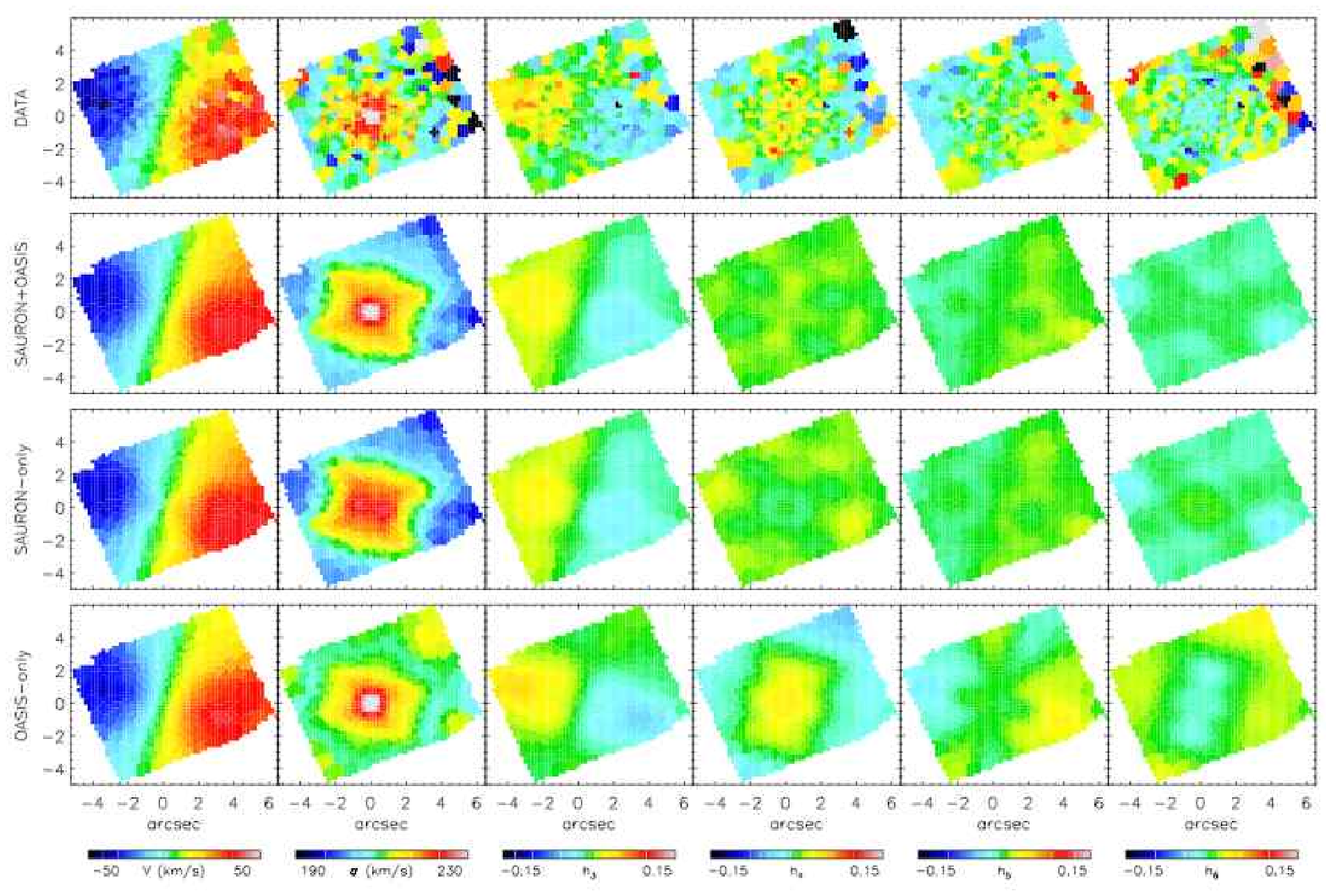}
	\caption{Analogous to Fig. \ref{SAUbest}, except here the third row from the top is a prediction of the {\tt OASIS} kinematics from the model fitting only {\tt SAURON} data, and the bottom row is the fit to the {\tt OASIS} kinematics from the model fitting only {\tt OASIS} data.  The model fitting only the {\tt SAURON} data clearly underpredicts the central velocity dispersion, due to the lack of high resolution kinematics in the vicinity of the black hole.  The model fitting only the {\tt OASIS} data, on the other hand, has such freedom with the orbital structure that it reproduces nearly all the features in the fields, including some of the noise.  Note that contour levels on the velocity dispersion have been rescaled from Fig. \ref{SAUbest} to aid the eye.}
	\label{OASbest}
\end{figure*}

The $\chi^2$ results from the edge-on stellar dynamical model are shown in the right panel of Fig. \ref{SAUandOASchi}, with the best-fitting model being described by a black hole of mass \sbhedgeerr\ and with M/L = \smlerr\ \ml\ (errors are 3$\sigma$).  The model predictions for the {\tt SAURON} and {\tt OASIS} kinematic fields are compared to the data in Figs. \ref{SAUbest} and \ref{OASbest} respectively; in general, these models do a very satisfactory job of reproducing the observations.

As a test of the effect of data resolution and spatial coverage on model results, we also ran models with the {\tt SAURON} and {\tt OASIS} data sets individually and show the resulting $\chi^2$ contour plots in Fig. \ref{SAUandOASchi}.  (We call these models the {\tt SAURON}-only model, which fits the {\tt SAURON} data and merely predicts the {\tt OASIS} kinematics, and the {\tt OASIS}-only model, which fits only the {\tt OASIS} data and predicts the {\tt SAURON} kinematics.)  Both of these supplementary models have a best-fitting black hole mass and M/L consistent with the model constructed using the combined data sets.

\begin{figure*}
	\includegraphics[width=18.3cm]{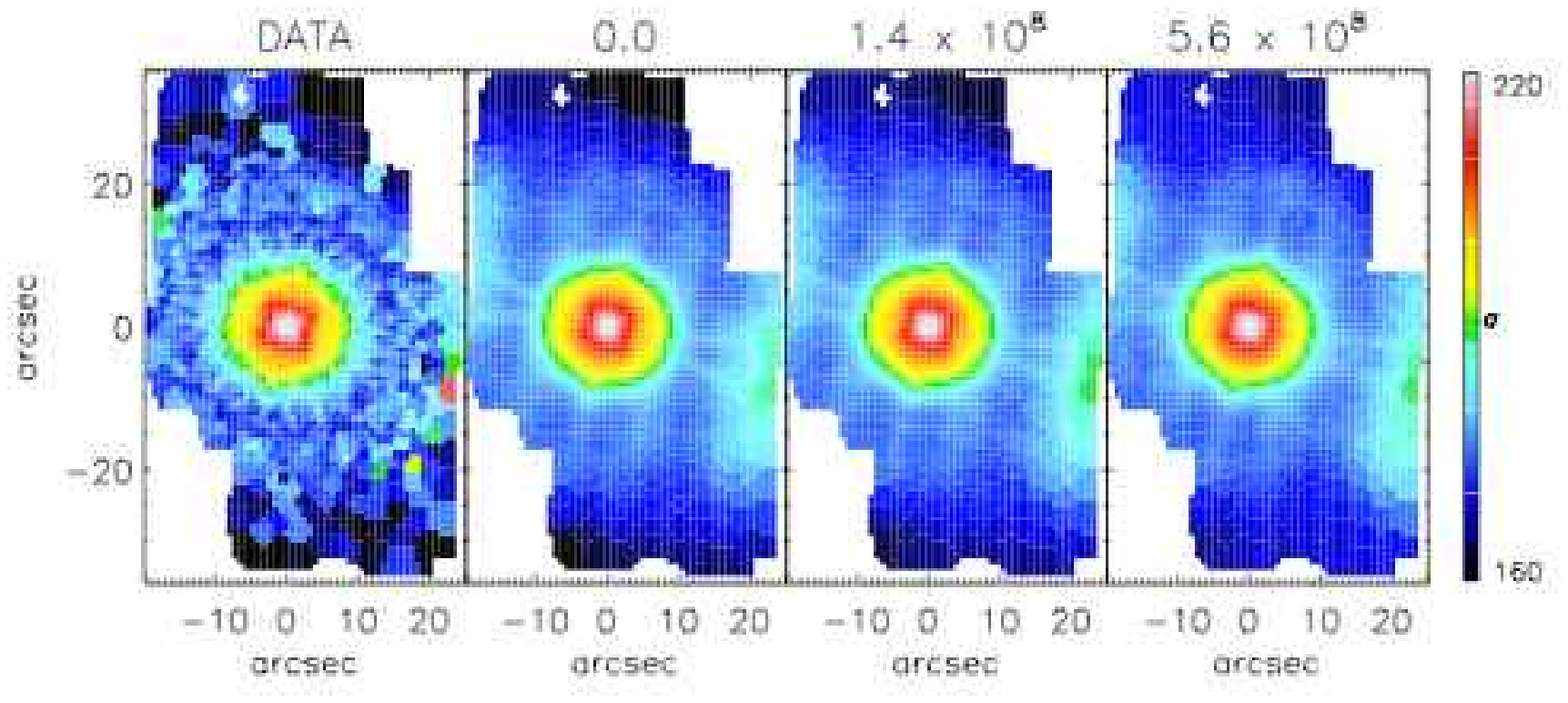}
	\caption{Velocity dispersions (in \kms) as observed in the {\tt SAURON} field (left) and as modelled with varying black hole masses (indicated above each model, measured in \msun).  Due to the small sphere of influence of any SMBH (on the scale of a {\tt SAURON} pixel in all cases), differences between the various models occur only in central pixel and are nearly undetectable.  For this reason, the {\tt SAURON} data alone are insufficient to constrain the mass of the black hole.  The models shown here are for the {\tt SAURON}+{\tt OASIS} data set; however, the results for the {\tt SAURON} datacube from the {\tt SAURON}-only model are virtually identical.  In both cases, the dispersion of the central pixel alone is sufficient to constrain the black hole mass.}
	\label{ssigmas}
\end{figure*}

\begin{figure*}
	\includegraphics[width=18cm]{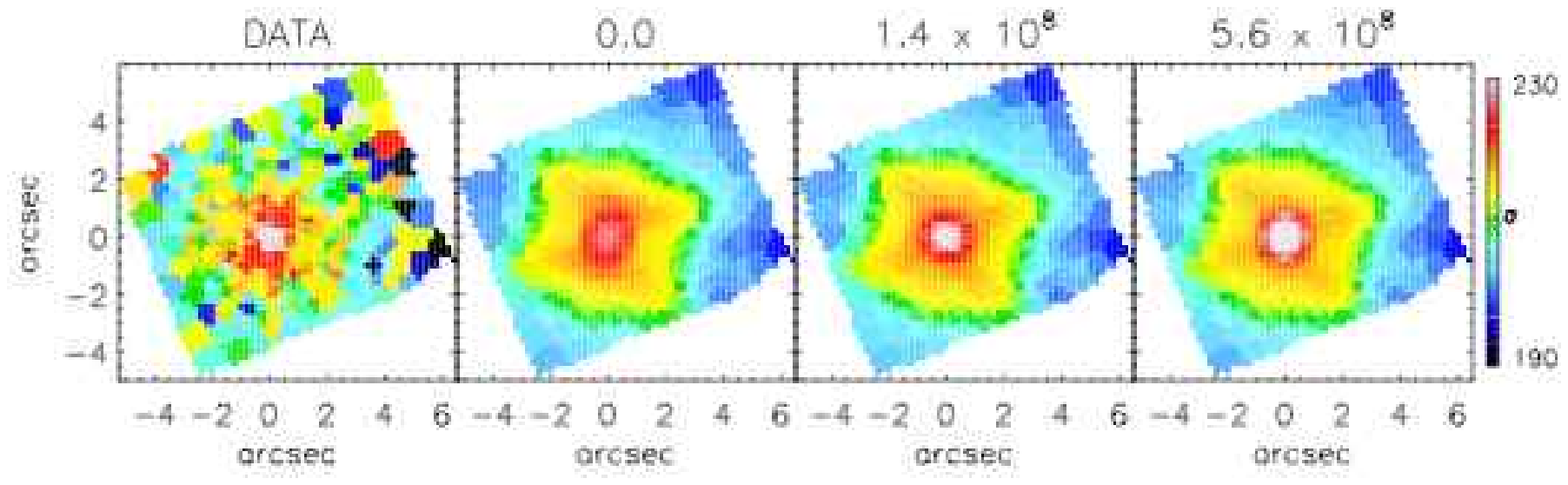}
	\caption{As in Fig. \ref{ssigmas}, but for the {\tt OASIS} field instead.  The high resolution {\tt OASIS} data probe the black hole's sphere of influence and thus provide significant leverage in measuring the mass of the SMBH.  The black hole mass is constrained almost entirely by the velocity dispersion (displayed here); this is demonstrated explicitly in Fig. \ref{SlitModel}.  The models shown reflect the {\tt SAURON}+{\tt OASIS} model; as is clear in Fig. \ref{SAUbest}, the {\tt OASIS}-only model is unable to properly constrain the orbital structure and is therefore useless in this sort of comparison.}
	\label{osigmas}
\end{figure*}

While it is clear that the wide-field {\tt SAURON} data play the dominant role in constraining the stellar M/L, it is rather remarkable that these data are also able to constrain the black hole mass so well, given that the sphere of influence $R_{\rm BH}$ $\sim$ \rbh\ is well below the resolution and seeing limit of the {\tt SAURON} observations.  This black hole mass measurement must therefore be due to minute differences in the model predictions for the centre-most spatial element.  This is apparent in Fig. \ref{ssigmas}, which shows the model velocity dispersions for different in black hole masses; the central regions of {\tt SAURON} field display almost no detectable difference in velocity dispersion for extremely large variations in black hole mass (and likewise for other moments of the velocity distribution).  This ability of the {\tt SAURON} data to discriminate between such similar models speaks to the high quality of the data (S/N$_{\rm center}$ $\approx$ 560), but care must be taken when further interpreting these results.  Figure \ref{OASbest} shows that the best-fitting {\tt SAURON}-only model drastically underpredicts the central velocity dispersion observed in the higher resolution {\tt OASIS} data.  Thus, while the {\tt SAURON}-only model is able to constrain the black hole mass, the lower spatial resolution of these data, and their correspondingly poor sampling of the black hole's sphere of influence, render them incapable of correctly measuring the central orbit structure.  This is even more clearly seen in Fig. \ref{osigmas}; the higher resolution {\tt OASIS} data, even in a model without a black hole, require a more physical (centrally-peaked) velocity dispersion distribution than that predicted for the {\tt OASIS} field by the {\tt SAURON}-only model.  This test is a simple example of both the strengths and limitations of using data of insufficient resolution to study the central regions of galaxies; with high-quality data it may be possible to obtain accurate black hole masses, but it would be inadvisable to use those results in a more detailed discussion of the stellar dynamics in the galaxy centre.

\begin{figure*}
	\includegraphics[width=18cm]{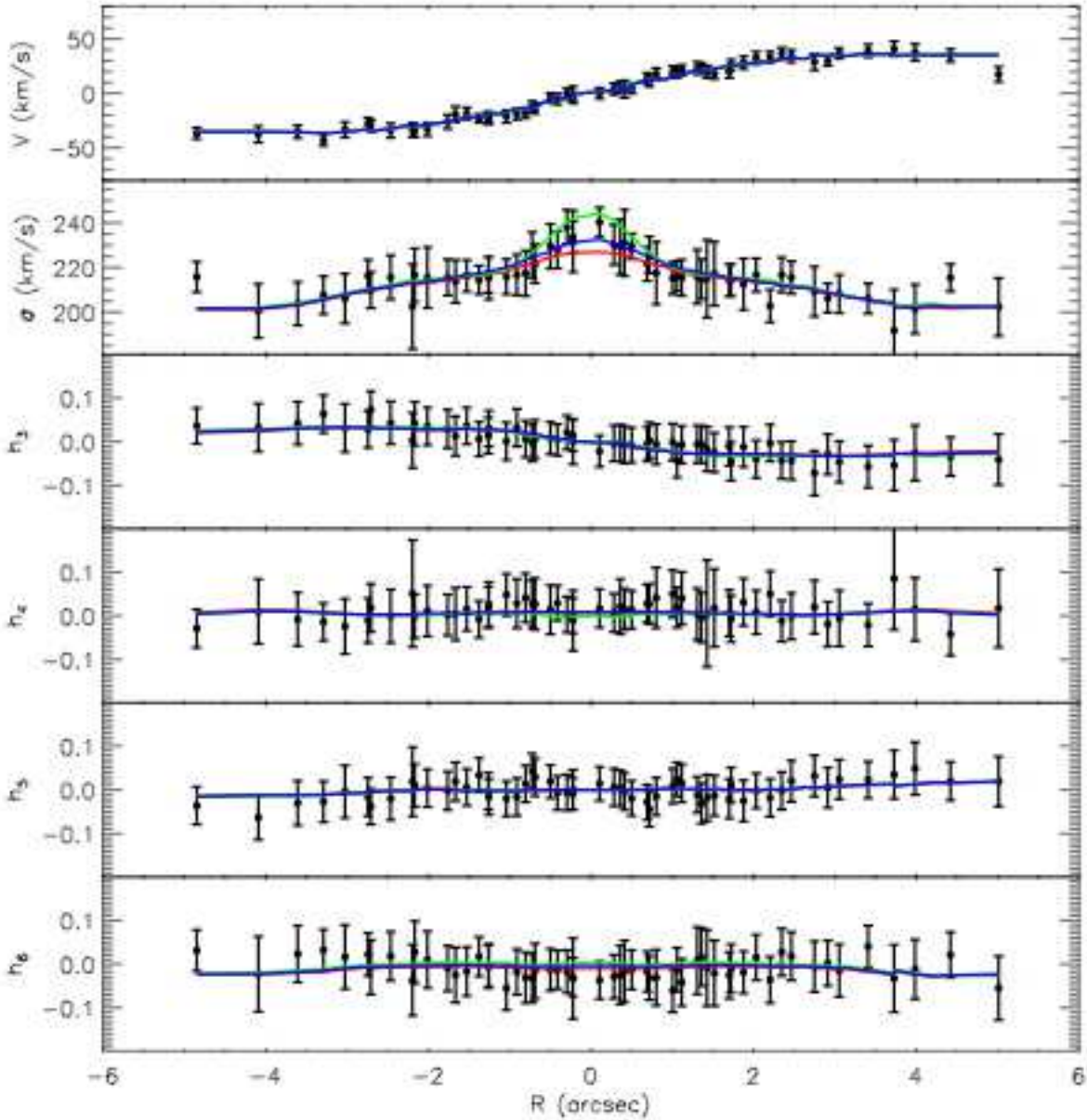}
	\caption{{\tt OASIS} kinematics along the major axis, compared to models of several SMBH masses: 0 (red), $1.4 \times 10^8$ \msun\ (blue), $5.6 \times 10^8$ \msun\ (green).  The velocity dispersions clearly provide the main constraint in measuring the black hole mass.  The models shown reflect the {\tt SAURON}+{\tt OASIS} model.  The model fits all the data points well within the error bars.  This is due to the plotted kinematics having been bi-symmetrised for input into the Schwarzschild model (see Section \ref{Schw3379}).  This reduces the scatter between adjacent points, but the error bars remain unaltered, since the number of data points is unchanged.  The unsymmetrised data are shown in Fig. \ref{SAUandOASkin}.}
	\label{SlitModel}
\end{figure*}

The {\tt OASIS} data, on the other hand, have higher spatial resolution than the {\tt SAURON} data but also have dramatically lower spatial coverage.  The ability of the {\tt OASIS} data to resolve the black hole's sphere of influence in two dimensions ensures that varying black hole masses will have noticeable effects on the predicted stellar kinematics.  At first, it is therefore a bit surprising that the $\chi^2$ contours for the {\tt OASIS}-only models (Fig. \ref{SAUandOASchi}) are so broad and that the {\tt OASIS} data alone are incapable of measuring the black hole mass with much significance.  Figure \ref{SAUbest} immediately resolves this issue; the lack of spatial coverage in the {\tt OASIS} data gives the model complete freedom to vary the orbital distribution throughout the galaxy, since it is constrained only in the central few arcseconds.  The prediction for the {\tt SAURON} (wide-field) data from the {\tt OASIS}-only models are utterly unrealistic, since models with so much freedom in the galaxy's orbital structure can not be depended upon to accurately extrapolate to large scales.  However, when the {\tt SAURON} data are included as constraints on the large-scale structure of the galaxy, the {\tt OASIS} data are a powerful probe of the black hole mass.  As shown in two dimensions in Fig. \ref{osigmas} and along the major axis with errors in Fig. \ref{SlitModel}, inspection alone is sufficient to locate the best-fitting model when the combined {\tt SAURON} and {\tt OASIS} data sets are used.  This suggests that the best-fitting model from the combined data sets also provides an accurate representation of the stellar orbit structure in the central regions of the galaxy.  To test this hypothesis, this best-fitting model was used to predict the higher spatial resolution STIS and FOS stellar kinematics, and these predictions were completely consistent with those measurements.  The ability of the {\tt OASIS} data to resolve the black hole sphere of influence therefore provides sufficient constraints on the central stellar orbit structure that the predicted stellar dynamics from this model can be extrapolated to much smaller scales.

Neither the results from the {\tt SAURON}-only model nor those of the {\tt OASIS}-only model are unexpected; similar results were seen by Copin, Cretton \& Emsellem (2004) in their models of NGC~3377 using {\tt SAURON} and {\tt OASIS} data for that galaxy.  They found, as here, that the wide-field {\tt SAURON} data are critical in constraining M/L; however, unlike this study, they also found that the {\tt SAURON} data alone were unable to provide useful limits on the SMBH mass in NGC~3377.  This difference is likely due to the combined effects of a slightly lower S/N in their {\tt SAURON} data and worse seeing during those observations, which conspire to render the NGC~3377 {\tt SAURON} data not quite sensitive enough to tiny differences in the predicted central velocity dispersions for varying black hole masses.  Their results with the {\tt OASIS} data, on the other hand, are identical to those here; taken alone, their {\tt OASIS} data were able to exclude the no-black hole scenario to high significance but otherwise provided few constraints.  They conclude that only the combined data set, with both high resolution data and wide-field data, is capable of determining both M/L and \mbh, as also seen here in Figs. \ref{SAUbest} and \ref{OASbest}.

Gebhardt et al. (2003) and Krajnovi\'{c} et al. (2005) have also tested the effects of spatial resolution and coverage, respectively.  Gebhardt et al. (2003) measured the black hole masses in 12 nearby elliptical galaxies with stellar dynamical models, using combined {\it HST} and ground-based data and also using ground-based data alone.  They found that the ground-based data yields a black hole mass consistent with that from the combined data set, but at a reduced significance.  Their high-resolution {\it HST} data were needed to probe the spheres of influence of the black holes, which mimics the need for {\tt OASIS} seen here.  They also note that models using {\it HST} data alone were unable to constrain the LOSVD at large scales, as seen here with the {\tt OASIS}-only model.  To study the effect of spatial coverage in more detail, Krajnovi\'{c} et al. (2005) constructed a series of models of a two-integral test galaxy with simulated {\tt SAURON} data out to 1 and 2 $R_e$.  Their results indicated that the orbital structure was recovered well by both models.  As the {\tt SAURON} data of NGC~3379 extends to roughly 1 $R_e$ (= 42\arcsec, Cappellari et al. 2006), the tests of Gebhardt et al. (2003) and Krajnovi\'{c} et al. (2005) illustrate that these wide-field data are both necessary and sufficient to measure the orbital structure of NGC~3379 on large scales.  When combined with the {\tt OASIS} data, which resolves the black hole's sphere of influence, the model can reliably measure the galaxy's orbital structure on both small and large scales.

Since there has been significant recent debate concerning the effects of regularisation on black hole mass measurements, we generated a set of models for the {\tt SAURON}+{\tt OASIS} data set without smoothing in integral space ($\Delta = 1 \times 10^3$).  The results are completely consistent with those of the $\Delta=4$ models; there is perfect agreement on the black hole mass, mass-to-light ratio, and velocity distribution (presented in the following section).  The uncertainties on these measurements are similar for the $\Delta=4$ and $\Delta=1\times10^3$ models, with those of the non-regularised models being slightly smaller.  This simple test thus demonstrates that including moderate regularisation of $\Delta=4$ in the models is not driving the results and consequently does not impact our conclusions (see also Section \ref{StellarIssues} for further discussion).

\begin{figure}
	\includegraphics[width=7cm]{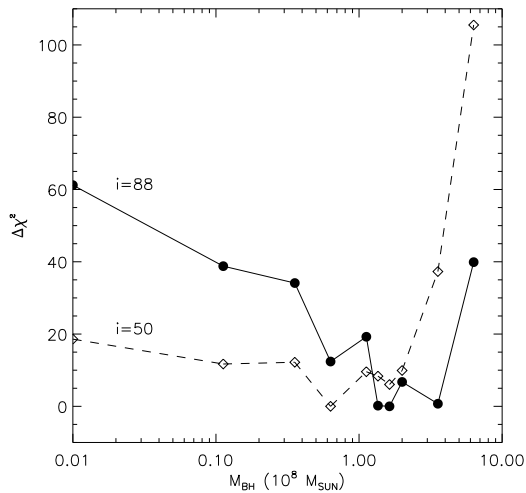}
	\caption{$\chi^2$ as a function of black hole mass, marginalised over M/L, for the two tested inclinations.  $10^6$ \msun\  has been added to the no-black hole model in order to include it on the log scale.  The derived best-fitting black hole masses vary by 50\% between the two inclinations.}
	\label{chibh}
\end{figure}

As a final test of the robustness of these results, we examined the effects of our assumed inclination on the measured black hole mass.  In addition to the edge-on models presented above, we generated a series of models near the minimum possible inclination.  Deprojection of the MGE model places a strict minimum of 35$^\circ$ on the inclination; however, inclinations of less than $\sim$ 45$^\circ$ require that the axial ratio of many of the Gaussian components be smaller than 0.2, implying that the galaxy is a thin disc.  We therefore adopt a test inclination of $i$=50$^\circ$, and the results are shown in Fig. \ref{chibh}.  Although the edge-on configuration is formally a better fit to the data, the tests of Krajnovi\'{c} et al. (2005) indicate that inclination determinations may be degenerate, and we restrict our interpretation of this result to noting that the assumed inclination plays a role in the measured black hole mass at the 50\% level.  Similar levels of dependence has been previously observed by e.g. Gebhardt et al. (2000b) for this galaxy and Verolme et al. (2002) for M32.  Because of the potential degeneracy in the determination of inclinations, we cannot claim a definitive measurement of the inclination of NGC~3379, and we therefore marginalise over the two tested inclinations in our error estimates.  This constrains the black hole mass to \sbhone\ (1$\sigma$), or \sbherr\ (3$\sigma$).  For our analysis of the model results, we use the best-fitting model, which has $i$=88$^\circ$, M/L=\sml, and \mbh=\sbh.

\subsection{Stellar Dynamics Near the SMBH}
\label{VelAnis}

The inclusion of the {\tt OASIS} high-resolution two-dimensional data in the stellar dynamical models provides us with the unique opportunity of probing the stellar velocity distribution in three dimensions in the vicinity of the central SMBH.  By measuring the kinematics of the solution set of orbits, we can study the full three-dimensional motions of stars.  In particular, we examine the velocity ellipsoid, the shape of which has been linked to the steepness of the central surface brightness profile (i.e., ``cusp" or ``core") and the nature of the central SMBH (i.e., single object or binary system, see Quinlan \& Hernquist 1997; Gebhardt et al. 2003).

\begin{figure}
	\includegraphics[width=8.5cm]{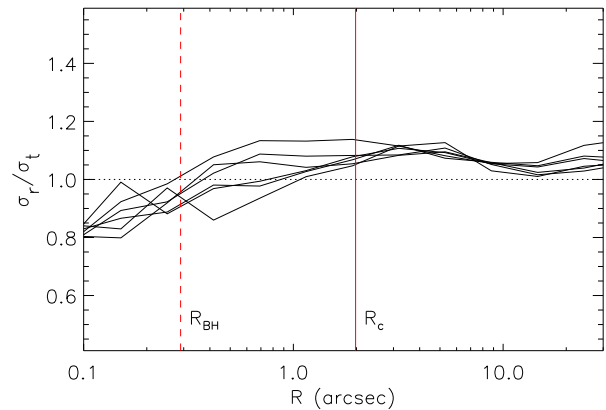}
	\caption{Velocity anisotropy of NGC~3379 as a function of radius along the angular sectors sampled in the Schwarzschild model.  The dashed line marks $R_{\rm BH}$ (=\rbh) and the solid line marks $R_c$ (=1\farcsec98).  The galaxy is characterised by mild radial anisotropy, tending towards isotropy within $R_c$.  (The trends shown here are also seen in the model with $\Delta=1\times10^3$ and in the model at $i$=50$^\circ$.)}
	\label{Butterfly}
\end{figure}

Figure \ref{Butterfly} shows the ratio of the radial velocity dispersion $\sigma_r$ of the stars to the their tangential velocity dispersion ($\sigma_t$, defined as $\sigma_t^2 = (\sigma_\phi^2 + \sigma_\theta^2)/2$) over all observed radii.  Note that $\sigma_\phi$ includes only random, and not ordered, motion such that an isotropic distribution is characterised by $\sigma_r$ = $\sigma_t$.  In NGC~3379, the anisotropy profile shows a mild radial anisotropy throughout most of the galaxy.  Within $R_c$, down to the limit of our spatial resolution, the distribution is consistent with isotropy.

However, in other core galaxies, the velocity distribution within $R_c$ has been observed to be tangentially biased (Gebhardt et al. 2003), and this correlation has been hypothesised to be the fossil remnant of a binary black hole in the galaxy centre.  In this picture, during the merger of two large galaxies with SMBHs, the black holes fall to the centre of the potential well through dynamical friction and form a binary.  Stars with radial orbits then pass close to this binary and are ejected at much higher velocities, allowing the binary to lose energy and shrink in size.  This process both reduces the number of stars in the centre-most regions of the galaxy, creating a lower-density ``core," and preferentially removes stars on radial orbits, creating a strong tangential bias in the velocity ellipsoid (Quinlan \& Hernquist 1997).

In detailed models of such systems, Milosavljevi\'{c} and Merritt (2001) have indeed shown that the formation and evolution of a hard black hole binary in a galaxy centre is accompanied by a change in the surface brightness profile from power-law to core and by a central tangential anisotropy whose spatial extent increases with time from $R_{\rm BH}$ to $R_c$ (see their Figs. 7 and 16).  As NGC~3379 has a very strong core in its surface brightness distribution, the results shown in Fig. \ref{Butterfly} make this galaxy not obviously explained by these simulations.  This discrepancy may be due to multiple merger events during the galaxy's history, which would then make it difficult to directly compare the details of the simulations and the observations (Milosavljevi\'{c} \& Merritt 2001).

\section{Gas Dynamical Modelling: Method}
\label{GasMod}

\subsection{Basics}
\label{Requirements}

To model the gas kinematics in the central gas disc of NGC~3379 and thus measure the mass of the central SMBH, we assume that a thin disc in circular motion is a reasonable description of the system.  This method has been commonly used to model nuclear gas discs (e.g. Macchetto et al. 1997; van der Marel \& van den Bosch 1998; Bertola et al. 1998; Barth et al. 2001), and it has been described extensively by those authors.

In the case of NGC~3379, Fig. \ref{WFPCbands} shows the gas disc major axis to be at PA $\sim$ 118$^\circ$, approximately 45$^\circ$ away from the rotation axis of the stars, a configuration that is seemingly inconsistent with the gas being on stable circular orbits.  However, both the MGE surface brightness parametrization and previous dynamical models (e.g. Statler 2001; Gebhardt et al. 2000b) suggest that NGC~3379 is intrinsically quite round, especially in the central regions, where the projected axial ratio q$\prime>0.9$.  In such a nearly spherical potential, there are no strongly preferred axes and therefore hardly any torques on a gas disc, regardless of its orientation.  Consequently, the disc in NGC~3379 is likely to be stable on a timescale much longer than its rotation period, allowing the gas to settle into circular orbits in a thin disc.

To model this disc, we can therefore use the thin-disc method of the aforementioned authors.  Briefly, from that assumption, it follows that the gas disc is rotating solely under the combined gravitational influence of the black hole and of a stellar potential.  The velocity contribution from the enclosed stellar mass is determined by using the MGE surface brightness model in combination with a constant {\it I}-band stellar mass-to-light ratio.  The resulting disc velocity field is then projected on to the plane of the sky by assuming an inclination $i_{\rm disc}$ of the disc.  The projected line-of-sight velocities of the disc are therefore parametrized by only three variables: \mbh, the stellar M/L, and $i_{\rm disc}$.

This ideal velocity field is then made more physical by accounting for gas velocity dispersions and for surface brightness variations across the field.  Since the mechanism responsible for the gas velocity dispersions is not understood, we fold a simple parametrization for the dispersions into the model (Section \ref{DispField}).  The resulting velocity field is then weighted by the \ha\ surface brightness, using the \han\  WFPC2 image (Section \ref{SB}).

We simulate observations of the disc through the STIS instrument to achieve the best possible match to the data.  Three slits are placed on the velocity field, with their locations described by a position angle relative to the major axis of the gas disc and by a spatial position corresponding to the galaxy centre.  There are thus six parameters in our model: \mbh, M/L, disc inclination, position angle of the slits, and $x$- and $y$-coordinates of the slit positions.  To measure the black hole mass in NGC~3379, we constrain these parameters as best as possible and then construct models with varying values of these parameters and find the best match to the data in the $\chi^2$ sense.

In practice, we used the procedures of Cappellari et al. (2002), which account for the finite STIS pixel size, instrumental broadening, and the STIS PSF.  A model PSF was constructed using the TinyTim software (Krist 1993) for a monochromatic source at 6600~\AA.  We then modified the method of Cappellari et al. (2002) to include the slit effect (the velocity shift induced by the non-zero width of the slit and its projection on to the STIS CCD) as in Barth et al. (2001), and also to use the WFPC2 \han\ image (rather than an analytic formula) as a description of the gas surface brightness.  The result of the model calculation is a simulated STIS observation of the emission lines in NGC~3379, with the model emission lines having the same spatial and spectral dimensions as the STIS data.  The kinematics of the model emission lines were extracted in exactly the same manner as was done on the actual STIS data, with single Gaussians being fit to the LOSVDs of individual rows.

\subsection{Intrinsic Velocity Dispersion}
\label{DispField}

The measured STIS (\nii\ $\lambda$6584) kinematics shown in Fig.~\ref{STISkin} display significant velocity dispersions, which peak at over 100~\kms\ in all three slits.  Preliminary models revealed that a black hole and instrumental effects were insufficient to account for such broadening in an unperturbed gas disc.  This phenomenon has been observed previously in similar studies in a number of galaxies (e.g. Verdoes Kleijn et al. 2002).  However, the responsible mechanism is currently not understood; the large line widths may result either from the non-gravitational motions of local turbulence, from a non-circular motion analogous to stellar asymmetric drift, or from unresolved rotation (e.g. van der Marel \& van den Bosch 1998; Verdoes Kleijn, van der Marel \& Noel-Storr 2004).

In the case of the velocity dispersions being due to turbulence, the dispersions have minimal effects on the model, which consists of only gravitational and instrumental considerations.  Under this assumption, we assign a constant (thermal) velocity dispersion of $\sigma$~=~20~\kms, which corresponds to a temperature of $\sim 5 \times 10^4$~K.

Several authors (e.g. Barth et al. 2001) have constructed models with a similar methodology to that described here, both with and without an ``asymmetric drift" term to account for the large dispersions.  Barth et al. (2001) find that the addition of asymmetric drift provides a much more satisfactory fit to their observed dispersions (which peak at $\sim$ 150\ \kms) and changes their best-fitting black hole mass by 12\% (Barth et al. 2001).  Since our results include uncertainties much larger than this (Section \ref{GasResults}), we do not include this term in our models, nor do we expect this omission to alter our results significantly.

In the case of NGC~3379, we find that the most satisfactory explanation of the high central dispersions is in fact that of unresolved rotation (see Section \ref{twist}).  In this case, a thermal velocity dispersion of $\sigma$ = 20 \kms\ is sufficient to explain much of the observed velocity dispersion structure.  However, due to the present limited understanding of gas velocity dispersions, a given model's ability (or lack thereof) to reproduce the observed velocity dispersions provides little insight into the accuracy of the model; for this reason, we do not include velocity dispersions when considering the goodness-of-fit of individual models.

\subsection{Surface Brightness}
\label{SB}

Since the STIS data are of the \ha\  and \nii\ $\lambda$6584 emission in the gas disc, a representation of the actual appearance of the gas disc at the \ha\  and \nii\  wavelengths is required.  The \han\  image (Section \ref{WFPC}) is a good first approximation; however, this image includes the WFPC2 PSF.  Since the surface brightness image (folded in with the model velocity field) will be passed through the simulated STIS optics, we ideally want an image of the gas disc without the influence of this PSF.

We created a model PSF for the F658N filter, generated at the pixel location of the galaxy centre, using the TinyTim software (Krist 1993).  Richardson-Lucy deconvolution was used to remove this PSF from the \han\  image.  Due to the low signal and diffuse nature of the image, the deconvolution was stopped after 10 iterations; further iterations amplified the noise to an unacceptable level.  The deconvolved \han\  image was then rotated to the same frame as the STIS data, using position angle information in the image headers.  The centre of the disc photometry is assumed to be spatially coincident with the location of the supermassive black hole (i.e., the centre of the model velocity field).

While both the \han\ surface brightness and velocity dispersion play significant roles in the model, they contain little or no information about the possible presence of a central SMBH.  Consequently, these quantities are omitted from the $\chi^2$ calculation.  To measure the black hole mass, we therefore computed $\chi^2$ for each model by comparing only the extracted model velocities (measured as described in Section \ref{Requirements}) with those of the data.

\section{Gas Dynamical Modelling: Results}
\label{GasResults}

\subsection{Unperturbed Models}
\label{BestFit}

Of the six parameters in the model (M$_{\rm BH}$, M/L, $i_{\rm disc}$, slit PA, and $x$- and $y$-locations of the slit), the values of four ($i_{\rm disc}$, slit PA, and $x$- and $y$-locations of the slit) have been previously determined, from the WFPC2 photometry (Section \ref{WFPC}), {\it HST} data headers, and STIS acquisition images (Section \ref{STIS}), respectively.  Preliminary models indicated that the STIS gas kinematics are consistent with these values, so we do not fit for them.  The STIS data provide little leverage in determining M/L, given their incomplete spatial coverage of the gas disc.  However, NGC~3379 has a well-measured M/L (\sml\ \ml\  in the {\it I}-band), both from the stellar dynamical model presented here as well as from other models (e.g. the Jeans model of Cappellari et al. 2006), and we adopt that value here.  This leaves only one parameter, the black hole mass, that must remain free in the gas dynamical models.  Models were therefore generated for a variety of black hole masses and compared to the data in the $\chi^2$ sense.

\begin{figure}
\begin{center}
	\includegraphics[width=6cm]{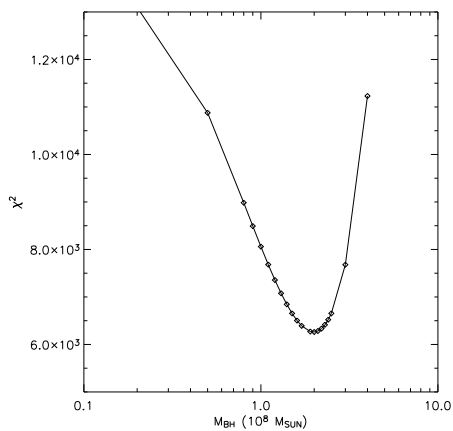}
	\caption{$\chi^2$ as a function of black hole mass for the thin disc gas dynamical model.  Points indicate individual models.  With $\sim$ 80 degrees of freedom, the reduced $\chi^2$ for even the best-fitting model deviates significantly from unity.  However, even this model is able to exclude the no-black hole scenario to over 6$\sigma$.}
	\label{BHnotwist}
\end{center}
\end{figure}

\begin{figure*}
	\includegraphics[width=16cm]{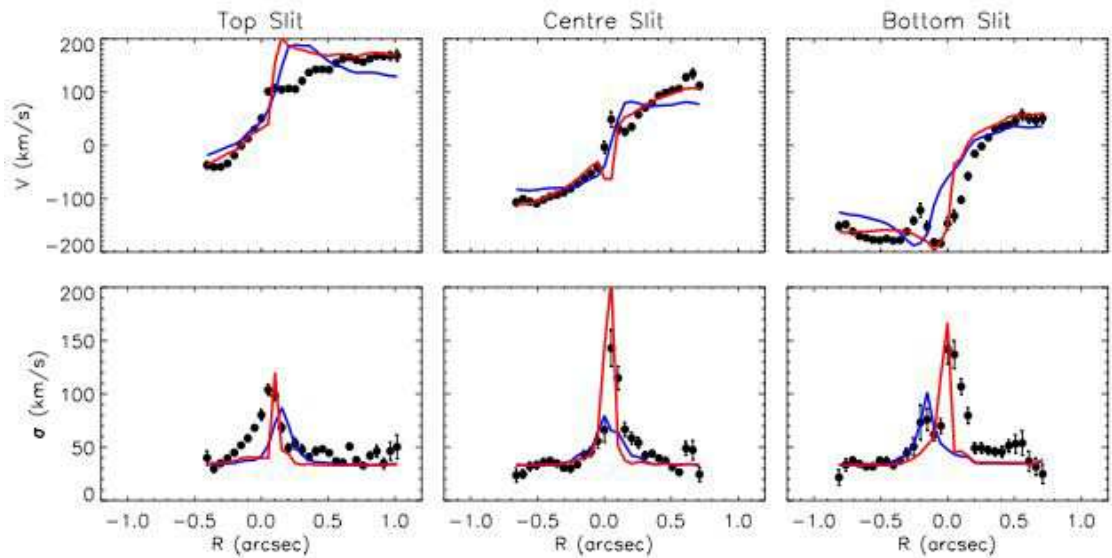}
	\caption{\nii\  velocities and dispersions of the central gas disc, as measured along all three slits (black), as predicted by the unperturbed model (blue), and as predicted by the twisted model (red).  One arcsecond at our assumed distance of 10.28 Mpc corresponds to 50 parsecs.}
	\label{gasLOSVD}
\end{figure*}

Figure \ref{BHnotwist} presents results from those models, with a clear $\chi^2$ minimum between $1 \times 10^8$ and $3 \times 10^8$ \msun.  Models lacking a central black hole are strongly excluded, as are those with a black hole of mass greater than $3 \times 10^8$ \msun.  The best-fitting model requires a black hole mass of \gbherr\ (3$\sigma$), and the emission lines corresponding to this fit are shown in Fig. \ref{BarthFig4}.  The gas velocities and dispersions measured from these lines are compared to the STIS gas kinematics in Fig. \ref{gasLOSVD}.

Despite the tight $\chi^2$ minimum, a visual comparison of the best-fitting unperturbed model to the data is rather unconvincing.  Most noticeable is the extreme overprediction of the velocity gradient at low radii in all three slits, which would seem to indicate a too-high black hole mass.  Indeed, the velocity gradient of the model fails to match that of the data at nearly all radii.  Furthermore, the model also slightly underpredicts the gas velocities at the edges of the disc (larger radii).

At the edges of the disc, the motion of the gas is mainly a reflection of the underlying stellar potential.   A possible solution to the data-model mismatch may therefore be found in lifting the assumption of a constant M/L throughout the galaxy.  In this case, a higher central M/L would increase gas velocities at the edges of the disc and simultaneously allow for a smaller black hole at the galaxy's centre.  Alternatively, the problem with the model may lie in the assumption of circular motion.  In the data of both the top and bottom slits, the peak velocity occurs at a radius of nearly zero (see Fig.~\ref{BarthFig4}).  Given that the slit position angles are offset from the major axis of the gas disc, any model of purely circular motion will result in the peak velocities in the top and bottom slit occurring at non-zero radii.  (This concept is elaborated on and illustrated in Section \ref{twist}.)  In the following sections, we draw on the combined constraints of the STIS and {\tt OASIS} gas kinematics to test both of these possible solutions.

\subsection{Fitting M/L}
\label{gasML}

The gas velocities are most sensitive to stellar M/L variations outside the sphere of influence of the black hole, a region that is only marginally probed by STIS due to the large position angle offset of the slits (see Fig. \ref{WFPCbands}).  The {\tt OASIS} data, on the other hand, provide a more comprehensive view of the gas kinematics over the entirety of the gas disc.  While stellar population studies suggest that there is no significant population gradient in the centre of NGC~3379 (e.g. McDermid et al. 2006), for completeness we examine the possibility of a higher M/L in the central regions of this galaxy.  To fit for the M/L, we use the {\tt OASIS} gas data, which has complete coverage of the gas disc and consequently samples the gas velocities further from the black hole.

For consistency, the {\tt OASIS} data are modelled as a series of slits, each of the same width as an {\tt OASIS} pixel.  The model is nearly identical to that used for the STIS data; the sole difference is the omission of the slit effect correction in the {\tt OASIS} model, since spatial and spectral stages in the IFU observations are decoupled.  Of the six free parameters, the disc inclination was measured from the WFPC2 photometry (as above), and the position angle of the ``slits" and their $x$- and $y$-locations can be accurately found from the IFU data.  {\tt OASIS} provides the capability of measuring the kinematic PA rather than merely the photometric one found using the WFPC2 image.  However, visual inspection of the data makes it immediately clear that the major and minor axes are not perpendicular (see Fig.~\ref{BarthOasis}).  Since the primary constraints on M/L come from the peak gas velocities at larger radii, we estimate the position angle of the gas disc from the major axis, despite the inconsistency with the minor axis.  The resulting position angle is offset from the photometric position angle by $-13^\circ$.  The central $x$- and $y$-locations are taken as the centre of the {\tt OASIS} velocity field.

\begin{figure}
\begin{center}
	\includegraphics[width=6cm]{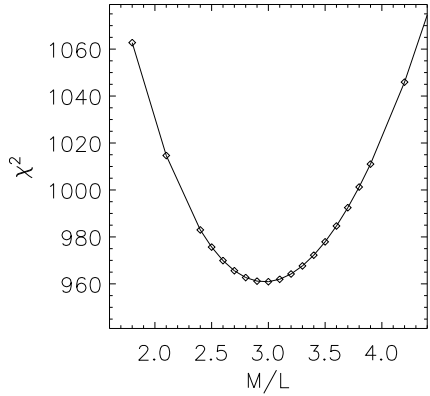}
	\caption{$\chi^2$ as a function of M/L, as measured from the {\tt OASIS} gas kinematics.  Points indicate individual models.  There are $\sim$ 150 degrees of freedom, making the $\chi^2_{\rm dof} \approx 6$ for the best-fitting model.}
	\label{MLnotwist}
\end{center}
\end{figure}

In principle, the remaining unknowns - \mbh\ and M/L - should both be varied in these models.  In practice, however, the large PSF of {\tt OASIS} renders the gas data insensitive to any velocity gradient in the vicinity of the black hole.  This need of the gas dynamical models, namely to detect Keplerian motion around the black hole in order to measure its mass, contrasts with the less stringent requirements of the stellar dynamical models, in which {\tt OASIS} measurements inside $R_{\rm BH}$ provide significant leverage on the central stellar orbital distribution and thus the black hole mass (as in Section \ref{StellarResults}).  Thus, while varying the black hole mass has a strong effect on the {\tt OASIS} stellar kinematic predictions, it has a negligible one on the {\tt OASIS} gas velocity predictions.  In the gas dynamical models of this data set, we therefore assume an \mbh\  of the unperturbed model's best-fitting value of \gbh\ and vary only the {\it I}-band M/L.

The gas velocities and dispersions were extracted at each position along each slit, as described in Section \ref{Requirements}, although the $\chi^2$ parameter was computed from the velocities alone.  Figure \ref{MLnotwist} shows how this parameter varies with the mass-to-light ratio, with a minimum at M/L = \gmlerr\ (3$\sigma$ errors).  The model corresponding to this fit is shown in Fig. \ref{BarthOasis}.  The {\tt OASIS} data very clearly constrain the central M/L to be almost exactly the value found in the stellar dynamical modelling; there is no evidence for a significant M/L gradient in the centre of NGC~3379.

\subsection{Possible Non-Circular Motion}
\label{twist}

With a higher central M/L clearly ruled out, the only method of accounting for the data-model mismatch in the unperturbed model is with some form of non-circular motion.  Unfortunately, there are numerous explanations for such motion (e.g. a bar, a spiral perturbation, a warp), and too few constraints on the type and shape of the perturbation.  Using any such perturbation model to make a black hole mass estimate is not feasible given our limited knowledge of the system.  However, it is possible to create a simple, though somewhat unphysical, model to determine whether the inclusion of non-circular motion can improve the model.  For this experiment, we assume our best-fitting black hole mass of \gbh\ and mass-to-light ratio of \sml.  (We note, however, that the following conclusions are unchanged when using the best-fitting stellar dynamical black hole mass of \sbh.)

Non-circular motion is introduced into the model described in Section \ref{GasMod} as a tilted ring model.  In the original model, the gas velocity at every radius was determined by the underlying potential (black hole plus stellar potential) and the disc inclination.  In the modified version, the underlying potential and inclination of the disc remain unchanged, but the position angle of the major axis at each radius is smoothly varied.  When the disc is projected on to the plane of the sky, this imposed twist in the major axis is visible as a twist in the velocity field of the disc.

\begin{figure}
	\includegraphics[width=8cm]{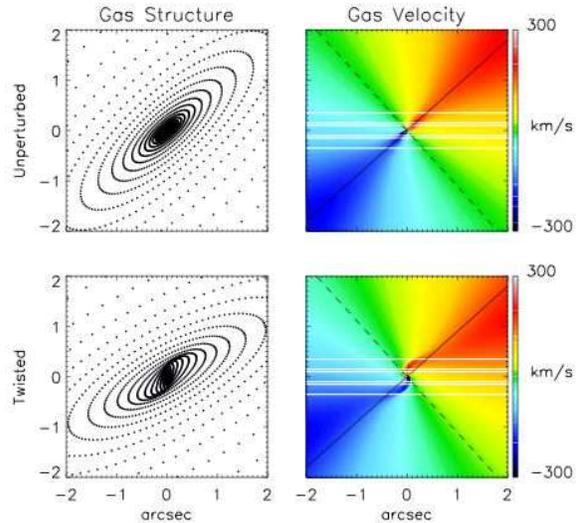}
	\caption{The gas structure (left), which loosely corresponds to density, and velocity field (right) in the unperturbed model (top) and in the twisted model (bottom), as projected on the plane of the sky.  The photometric major and minor axes of the gas disc are indicated by the solid and dashed black lines respectively.  It is evident that a non-twisted model requires that the peak velocity in the top and bottom STIS slits (white) must occur away from $r'$=0, whereas the twisted model permits the peak velocity to occur at $r'$=0.  The latter scenario is in keeping with the observations (Fig. \ref{BarthFig4}).  For clarity, these plots have been aligned with the same position angle as the WFPC2 images in Fig. \ref{WFPCbands} such that the STIS slits are horizontal.}
	\label{VfieldTwist}
\end{figure}

The shape and extent of this artificial twist is constrained by the regularity of the {\tt OASIS} velocity field at large radii (Fig. \ref{BarthOasis}).  The gas kinematics of the centre STIS slit (Fig. \ref{STISkin}), on the other hand, require a significant deviation from the photometric major axis in the centre of the disc.  We therefore impose a twist of the form

\[
\theta_{\rm offset} = A {\rm e}^{-r/B} + C ,
\]

\noindent
where $A$ gives the intensity of the twist, $B$ gives its radial extent, and $C$ allows for a position angle offset between the kinematic and photometric major axes, as seen in the {\tt OASIS} data.  Given the arbitrary nature of this model, a full $\chi^2$ minimization was not performed over these parameters; instead they were varied by hand until the model reproduced most of the features in the STIS data.  The final model was parametrized by $A$=2.4, $B$=0.25\arcsec, and $C$=-13$^\circ$ (see Fig.~\ref{VfieldTwist}).

The simulated STIS emission lines and LOSVD moments for this model are shown in Fig. \ref{BarthFig4} and Fig. \ref{gasLOSVD} respectively.  It is immediately apparent that many of the problematic features of the STIS gas velocities are reproduced quite well with this ad hoc model.  This is reflected in the $\chi^2$ for the STIS data against the twisted model, which is approximately half that of the best-fitting unperturbed model of Section \ref{BestFit}.  It is interesting to note that the gas velocity dispersions observed with STIS are generally predicted more accurately with this model, suggesting that the high dispersions are in fact unresolved rotation.  However, the unphysical nature of this twist and the lack of understanding of these dispersions prevent this correspondence from leading to any strong conclusions.

As a consistency check, this twisted model was also applied to the {\tt OASIS} gas model, and the results are presented in Fig. \ref{BarthOasis}.  Although the blurring effects of the {\tt OASIS} PSF remove most of the evidence of the twist, it is clear that the twisted model is an equally valid representation of the data, and indeed the $\chi^2$ of the twisted model is slightly lower than that of the unperturbed one.

The velocity field and density structure corresponding to this twisted model are shown in Fig. \ref{VfieldTwist}.  It is encouraging that, although arbitrary, the model that matches the gas kinematics produces a spiral-like structure that is hinted at in the WFPC2 photometry (Fig.~\ref{WFPCbands}).  At the level of both the photometry and kinematics, this simple twisted model is an adequate representation of the gas disc in NGC~3379 and certainly a better representation than one restricted to perfect circular motion.

\section{Discussion}
\label{Discussion}

\subsection{The Black Hole in NGC 3379}

Our stellar dynamical models, constructed using the combined {\tt SAURON}+{\tt OASIS} data set, yielded a black hole mass estimate of \sbherr\ (3$\sigma$), or \sbhone\ (1$\sigma$).  This is similar to the black hole masses found previously in other stellar dynamical models.  Using two-integral models and ground-based long-slit data, Magorrian et al. (1998) derived a black hole mass of $(3.9 \pm 0.4) \times 10^8$ \msun\ (1$\sigma$).  Gebhardt et al. (2000b) improved upon this method with the inclusion of {\it HST}/FOS data and three-integral models to measure a black hole mass of $1^{+1.0}_{-0.4} \times 10^8$ \msun\ (1$\sigma$), completely consistent with the results derived here.  Gebhardt et al. (2000b) attribute the smaller black hole in the three-integral model to a mild large-scale radial anisotropy predicted by their models (and seen here as well) that two-integral models are unable to reproduce.  This very good agreement between the Gebhardt et al. (2000b) three-integral models and those presented here, in both black hole mass and in general trends in orbital anisotropy, is also a nice check of the robustness of black hole mass determination using different data sets and different model implementations.

However, all of these stellar dynamical models employ nearly identical methods of measuring black hole masses.  A more objective comparison can be made using the results of completely different techniques.  Unfortunately, our gas dynamical results are not conducive to strong conclusions.  The unperturbed model, even at the best-fitting black hole mass of \gbh, is unable to reproduce most of the features of the observed data.  The twisted model produces a much more satisfactory result, fitting both the perturbed velocity field and the density enhancements visible in the \han\ image.  However, there are insufficient constraints on the true nature of this perturbation to generate a physical model and to perform a fit for the mass of the black hole.  We therefore only conclude that the gas kinematics are consistent with the black hole mass measured in the stellar dynamical models.

We are also able to study the region around this black hole in great detail with the stellar dynamical results.  Much of the galaxy is characterised by a mild radial anisotropy in the orbital structure, which becomes consistent with isotropy within $R_c$.  This result is rather inconsistent with current simulations, which suggest that the galaxy's strong core signature should be accompanied by a tangential anisotropy within $R_c$.  These surprising results have also been found in M87, a similarly round elliptical with a strong core, by Cappellari \& McDermid (2005).  Using integral-field data and three-integral models of M87, they measure a slightly radial velocity distribution across the galaxy and down to 0.5$R_{\rm BH}$; within that radius, they marginally resolve a shift towards tangential anisotropy.  Both M87 and NGC~3379 also have an $R_c$ approximately three times larger than that predicted by the simulations of Milosavljevi\'{c} \& Merritt (2001), suggesting that their formation histories include multiple merger events, which may be complicating this comparison of the data to single-merger simulations.

\subsection{Reliability of Stellar Dynamical Models}
\label{StellarIssues}

The stellar models described here have two main sources of uncertainty: the data themselves and the model assumptions.  As seen in Section \ref{StellarKin}, kinematic data are always subject to such systematic effects as template mismatch, in addition to statistical errors.  In this sense, NGC~3379 is an ideal galaxy for stellar dynamical modelling, since the wealth of kinematic data available in the literature and in the {\tt SAURON} and {\tt OASIS} data sets provides a good consistency check on all kinematic measurements.

When modelling this data, we have implicitly assumed that NGC~3379 can be well-described as an axisymmetric system, and that the MGE parametrization's lower limit on inclination excludes face-on models.  However, the true shape of NGC~3379 has been the subject of much speculation and, indeed, remains in doubt.  On the basis of photometric and kinematic similarities of this galaxy to NGC~3115, Cappaccioli et al. (1991) and Statler \& Smecker-Hane (1999) have proposed that NGC~3379 is in fact a nearly face-on S0 galaxy.  Statler (2001) modelled the shape and orientation of this galaxy using ground-based long-slit data and generally confirmed this result.  Their best-fitting model is a nearly axisymmetric galaxy (T = 0.08)  seen at moderate inclination ($i \approx 40^\circ$), although they were not able to completely exclude highly triaxial models.

From the {\tt SAURON} stellar kinematic observations, we are able to place additional constraints on the shape of NGC~3379 and make a strong qualitative argument for axisymmetry.  The integral field data have no minor axis rotation and have well-aligned kinematics and photometry, with a kinematic position angle of $72^\circ \pm2^\circ$ (found with the kinemetry method of Krajnovi\'{c} et al. 2006) and a photometric position angle of $68^\circ \pm2^\circ$ (from the MGE model).  More generally, NGC~3379 shows significant rotation, and all similar galaxies in the {\tt SAURON} survey have well-aligned kinematic and photometric axes and therefore likely do not deviate significantly from axisymmetry (Cappellari et al. 2005).

The combined results of Statler (2001) and the {\tt SAURON} observations effectively rule out highly triaxial models, making the most probable description of NGC~3379 that of a nearly axisymmetric galaxy, seen at a moderate inclination.  To test the effects of this possible configuration, we generated a series of models with $i$=50$^\circ$ and found that the best-fitting black hole mass differed from that of the edge-on models by 50\%.  While we include this effect in our final error estimations, we do not expect that the relaxation of our assumptions of axisymmetry and an 88$^\circ$ inclination will significantly affect our conclusions.

Finally, it has been suggested by e.g. Valluri et al. (2004) that there are fundamental problems with the reliability of three-integral dynamical modelling.  They find black hole masses to be dependent on (extreme) variations of the regularisation parameter $\Delta$, which biases the orbital weights towards a smoothly-varying integral space.  Verolme et al. (2002) explicitly tested the effects of varying $\Delta$ from 0.1 (high regularisation) to $\infty$ (no regularisation).  They found that the optimal balance between fitting the observables and enforcing smoothness in integral space occurred at $\Delta=4$.  In our own tests of our implementation of the Schwarzschild method, we have likewise found that $\Delta=4$ is most effective at recovering input parameters, including black hole masses.  Cretton \& Emsellem (2004) have shown that it is the absence of regularisation that leads to significant problems in black hole mass measurement, in that orbital weights vary rapidly (and unphysically) as a function of the integrals of motion.  We have tested this explicitly for NGC~3379 by running a set of models without regularisation and found no effect on our measured black hole mass or on the derived three-dimensional stellar velocity distribution.  We therefore adopt a moderate regularisation of $\Delta=4$ in our models and do not expect this issue to affect our results.

However, perhaps the most convincing argument for the robustness of the stellar dynamical black hole mass determination is found in Fig. \ref{osigmas}, in which the best-fitting model is immediately evident through inspection of the {\tt OASIS} data and models.  This capability is unique to models with integral field data, as becomes clear in a comparison with the previous models of this galaxy by Gebhardt et al. (2000b).  Using ground-based long-slit data plus an FOS spectrum, they were able to use $\chi^2$ to differentiate between models with varying black hole masses.  However, a visual comparison of the model with no black hole and that with a black hole of mass $1 \times 10^8$ \msun\ shows almost no discernible difference between the two (see their Fig. 11).  The combined {\tt SAURON}+{\tt OASIS} integral-field data, on the other hand, provides such strong constraints on the orbital distribution that the models are unable to reproduce the data with anything other than a black hole of mass $\sim$ \sbh\ (Fig. \ref{osigmas}).

\subsection{Reliability of Gas Dynamical Models}

The orientation of the nuclear gas disc in NGC~3379, which lies $\sim 45^\circ$ away from the rotation axis of the stars, has also played a role in discussions of the intrinsic shape of the galaxy.  In order for this configuration to be completely stabilised, there must either be a small degree of triaxiality in the galaxy or NGC~3379 must be nearly face-on, such that the gas disc is in an orbit close to polar.  In either case, as in the case of axisymmetry, Statler (2001) demonstrated that the disc must be decoupled from the main body of the galaxy (with possible origins from intergalactic H{\small I}, see e.g. Schneider 1989).  However, the models of Statler (2001), as well as the MGE surface brightness parametrization of NGC~3379, reveal that the centre of this galaxy is intrinsically quite round.  There are thus no strong torques on the gas disc, making its orientation at least marginally stable.  Nonetheless, given the position angle offset of this disc, it is perhaps unsurprising that the gas is not moving in perfect circular motion and that its distribution may in fact be (slowly) evolving.

Many elliptical galaxies are known to exhibit kinematic misalignments between the nuclear dust and stars, and it may be that the gas and dust discs in such objects are not generally settled into Keplerian orbits around a central black hole (e.g. van Dokkum \& Franx 1995; Emsellem, Goudfrooij \& Ferruit 2003).  However, even in the case of aligned stellar and gas kinematics, Verdoes Kleijn et al. (2002) were still unable to reconcile their gas and stellar dynamical measurements of the black hole mass in NGC~4335.  In that case, as here, the nuclear gas appears regular, and yet its high central velocity dispersions are better described with the inclusion of non-circular motions in the model.  Recent simulations have shown that spiral-like perturbations, whose unresolved rotation can cause such high velocity dispersions, can be generated by a central black hole binary (Etherington \& Maciejewski 2006); unfortunately, this scenario is not obviously consistent with our analysis of the stellar orbital structure.  It therefore remains to be seen whether the ``twisted" phenomenon observed in the gas disc of NGC~3379 is common in early-type galaxies and, if so, what causes this morphology.

Without the introduction of non-circular motion, inspection of the gas kinematics along the central slit of NGC~3379 suggests a smaller black hole mass.  Slight deviations from circular motion in a gas disc may therefore cause Keplerian models to underestimate the mass of the central black hole.  Indeed, in all of the existing comparisons of gas and stellar methods to date, the gas dynamical mass measurements are smaller than those derived from the stars.  This may be indicative that thin disc, circular models are too simplistic to describe most gas distributions in the centres of elliptical galaxies, although the limited number of gas-star comparisons is not sufficient for drawing such strong conclusions.

\section{Conclusions}
\label{Conclu}

As a nearby and seemingly normal elliptical galaxy, NGC~3379 seems, at face value, an ideal candidate for a comparison of stellar and gas dynamical black hole mass measurements.  We have used the most extensive combined stellar data set available for this galaxy, with the most detailed two-dimensional probe (3 $\times$ 3 pixels in {\tt OASIS}) into the region of influence of its black hole.  From this data, we measured a black hole of mass \sbherr\ (3$\sigma$), which agrees with results from previous stellar dynamical models based on longslit data and with the predictions of the \bhs\ relation.

NGC~3379 also has what appears to be a regular central gas disc, which is photometrically well-defined and has clear rotation.  Although it is not in the equatorial plane of the galaxy, the nearly spherical shape of the core allows the disc to be long-lived.  This paper presents the first detailed investigation into the kinematics of this disc, with the first dynamical model of it.  However, despite the apparent regularity of the gas structure and kinematics, the disc in fact displays significant deviations from circular motion.  Our nominal estimate of the black hole mass from the gas kinematics is \gbherr\ (3$\sigma$), although the corresponding model fits the data quite poorly.  Given the lack of constraints on the type and shape of the kinematic perturbation, it is not possible to properly model the true dynamics of the disc; our perturbed model provides a very satisfactory fit to the data but can only conclude that the disc is irregular and cannot provide an independent black hole mass measurement.

This paper's comparison of black hole masses derived using both stellar and gas dynamical methods is one of a very small number of such comparisons; however, to date all of these have revealed limitations in the data and in the model assumptions that have complicated and biased the black hole mass measurements.  There has yet to be a comparison of gas and stellar dynamical methods that shows the two methods to be in unequivocal agreement.  Cappellari et al. (2002) and Verdoes Kleijn et al. (2002) both found that gas and stellar dynamical black hole mass estimates can differ on the order of factors of 10, and in NGC~3379, the results are somewhat inconclusive.  It is clear that in these three cases the black hole masses cannot be trusted to level of the formal errors, but rather that systematic effects in the models tend to dominate the uncertainties.  

This should serve as a caution, since our present understanding of central SMBHs comes largely from black hole masses that have been measured using only a single method.  In particular, all of the very low mass SMBHs have been measured using stellar dynamical techniques, while most of the very high mass SMBHs have been studied via the surrounding nuclear gas.  While recent results have made it apparent that we have much to learn about galaxy evolution {\it from} \bhs, the results presented here indicate that we still have much to learn {\it about} \bhs\ as well.

\section*{Acknowledgments}

It is a pleasure to thank the entire {\tt SAURON} team for their efforts in carrying out the {\tt SAURON} observations and data preparation, for many fruitful discussions, and for a careful reading of this manuscript.  This paper was also improved considerably by helpful comments from the anonymous referee.  We are grateful to Eric Emsellem and Pierre Ferruit for taking the {\tt OASIS} observations and to the CFHT staff for their support of those observations.  We also wish to thank Patricia S\'anchez-Bl\'azquez and Reynier Peletier for early access to their MILES stellar template library.  The {\tt SAURON} project is made possible through grants 614.000.301 and 614.031.015 from ASTRON/NWO and financial contributions from the Institut National des Sciences de l'Univers, the Universit\'{e} Claude Bernard Lyon I, the Universities of Durham, Leiden, and Oxford, the British Council, PPARC grant `Extragalactic Astronomy \& Cosmology at Durham 1998--2002,' and the Netherlands Research School for Astronomy NOVA.  KLS is grateful for the repeated hospitality of the Sterrewacht Leiden, as well as that of the CRAL-Observatoire and of the University of Texas at Austin.  MC acknowledges support from a VENI award 639.041.203 awarded by the Netherlands Organization for Scientific Research (NWO).  KG and TSS received support through NASA grant GO-08589 from the Space Telescope Science Institute, which is operated by the Association of Universities for Research in Astronomy, Inc., under NASA contract NAS5-26555.

\bsp

\label{lastpage}

\end{document}